TALLINN UNIVERSITY OF TECHNOLOGY
School of Information Technologies

Gamaralalage Hiruni Sachinthana Gunaratne 184724IVGM

# IDENTIFYING BARRIERS IN E-INVOICING PROCESS TO INCREASE EFFICIENCY AND RAISE THE LEVEL OF AUTOMATISATION OF WORKFLOWS

Master's Thesis

Supervisor: Ingrid Pappel
PhD

Tallinn 2020

TALLINNA TEHNIKAÜLIKOOL
Infotehnoloogia teaduskond

Gamaralalage Hiruni Sachinthana Gunaratne 184724IVGM

# PEAMISTE PROBLEEMIDE TUVASTAMINE E-ARVETE MENETLUSPROTSESSIS SUURENDAMAKS TÕHUSUST TÖÖVOOGUDE AUTOMATISEERITUSE TASEME TÕSTMISE ABIL

Magistritöö

Juhendaja: Ingrid Pappel
PhD

Tallinn 2020

## Author's declaration of originality

I hereby certify that I am the sole author of this thesis. All the used materials, references to the literature and the work of others have been referred to. This thesis has not been presented for examination anywhere else.

Author: Gamaralalage Hiruni Sachinthana Gunaratne

07.05.2020

# Abstract

E-invoicing is a fast-budding e-service in Europe as well as in the world which is identified as a substantially significant element in progressing towards the goals of 'Digital Economy' in the European Union.

This Master's thesis focuses on identifying inefficiencies in e-invoicing systems currently in use and the opportunities to apply emerging technologies such as artificial intelligence, robotic process automation and blockchain to increase efficiency and level of automatization. The study incorporates expert opinions and users' perceptions in e-invoicing systems on necessities for automation. The scope is e-invoicing systems in the Baltic region which includes Estonia, Latvia and Lithuania.

To conduct the analysis, interviews with experts from the Baltic region and a survey for the e-invoicing users is incorporated. The study is qualitative research.

Based on the analysis, it emerged that drawbacks in e-invoicing systems can be identified as operational, information security-related and technological. The automation level of e-invoicing systems is moderate and there is room for improvement. The user's perceptions towards automation of the systems are positive. The functionalities which can be improved with emerging technologies are discovered by the author and the advantages of using emerging technologies in the context are explained. A software ecosystem designed by the author is presented as the recommendation.

This thesis is written in English and is 89 pages long, including 6 chapters, 22 figures and 9 tables.

# List of abbreviations and terms

| | |
|---|---|
| EDI | *Electronic Data Interchange* |
| e-invoice | *Electronic Invoice* |
| e-governance | *Electronic governance* |
| SME | *Small and Medium-sized Enterprises* |
| e-commerce | *Commercial transactions conducted electronically on the Internet* |
| e-service | *Electronic Service* |
| PEPPOL | *Pan-European Public Procurement On-Line* |
| PDF | *Portable Document Format* |
| EU | *European Union* |
| B2B | *Business to business* |
| B2G | *Business to government* |
| ERP System | *Enterprise Resource Planning System* |
| CEF | *Connecting Europe Facility* |
| e-Delivery | *Electronic Delivery* |
| e-State | *Electronic State (usually to address a country or a member state who has a lot of e-services)* |
| RPA | *Robotic Process Automation* |
| AI | *Artificial Intelligence* |
| SCF | *Supply Chain Finance* |
| BCT | *Blockchain Technology* |
| OCR | *Optical Character Reading* |
| ITL | *Estonian Association of Information Technology and Telecommunications* |
| FMCG | *Fast-moving consumer goods* |
| HoReCa | *Hotel/Restaurant/Catering* |
| IoT | *Internet of Things* |

# Table of contents

# List of figures

# List of tables

# 1 Introduction

## 1.1 Background

### 1.1.1 The Interest in e-Invoicing

Electronic invoicing is defined as "an invoice that has been issued, transmitted and received in a structured electronic format which allows for its automatic and electronic processing" [1]. According to Salmony and Bo Harald, "the e-invoice is a pivotal document in the supply chain whose digitalisation will generate savings in its own right, as well as contributing to many other benefits along the supply chain"[2]. It is said to be that the e-invoices should consist of fully structured data, therefore it can be automatically processed [2].

E-invoicing was known to be growing rapidly in the world with over 400 e-invoicing service providers active in Europe, transacting 3.3 billion Euros worth of e-invoices globally in 2017. And it is predicted to be 16.1 billion Euros in 2024[5]. According to the Digital Economy and Society Index Report 2019, not only the large corporations but also the SMEs in the EU are keener on automating their business processes. As an example, in 2018 23% of the SMEs and 47% of the large corporations in the European Union are sending e-invoices suitable for automated processing, which is a significant number [8]. Given that e-invoicing is one of the main e-services in e-commerce and e-governance [9], the thesis work revolves around e-invoicing.

In 2014, a directive on electronic invoicing in public procurement was voted by the European Parliament and Council which is known as DIRECTIVE 2014/55/EU. This directive emphasises on defining a common standard for e-invoicing which will develop interoperability within the European Union [1]. According to the directive, the European standard on e-invoicing stipulates invoices in a consistent organised format which offers an automatic reading of e-invoices into computer systems. The key target is to process the invoice automatically, and visualization of human-readable format is complimentary. As the bigger picture of the above directive, the European Union aims to create a 'digital single market' with no online barriers to flows of goods, services and data with the Europe 2020 ten-year plan [5]. According to the European Union's six strategic priorities for 2019-2024, 'A European Green Deal' ( Striving to be the first climate-neutral continent ) and 'A Europe fit for the digital age' (Empowering people with a new generation of technologies) [6] directly connects with the usage of e-invoicing within the member states.

e-Invoice is an essential document which creates advantages of its own in addition to the possibility of digitization of business processes in the supply chain which endorses efficiency and effectiveness [15]. According to Salmony and Harold, e-invoicing is a prospect to improve the competitiveness between the trading companies, increase customer satisfaction, faster payments, enhance the movement of cash and lessen credit-losses. Resting on all the advantages, e-invoicing has direct involvement in saving the environment [2].

**1.1.2 Why the Baltic Region?**

Estonia being a mature digital society with a well-functioning e-governance ecosystem [3], it is significant that the country's economic relations with the other countries also should head towards a digital arrangement. Estonia has been a member-state of the European Union since 1st May 2004 and is obliged to adhere to the directives imposed by the European Union [4].

The smallest unit that Estonia belongs to is the Baltic region. The Baltic countries which are Estonia, Latvia and Lithuania share common features and history similar to each other. According to Aurélien Poissonnier, a quantitative analysis of the economy of three Baltic countries reveals that there is an integrated Baltic economy which consists of a sizable part by common factors and economic links [7]. And also, a lot of businesses are operated within the Baltic region and have business relations among these three countries. Hence the author of the thesis decided to scope the research within the Baltic region because study results can be generalized to the whole region along with comparing some of the results within each country in the Baltics.

### 1.1.3 The Interest in Emerging Technologies

Emerging technologies such as Artificial Intelligence (AI), Internet of things (IoT), Robotic Process Automation (RPA) and Blockchain hold great significance when they are amalgamated with specific systems. Digital revolution is taking place globally enhancing the global markets concerning industries. Business sectors such as industrial manufacturing, healthcare, financial services, food services, automotive etc are progressively adopting Al, IoT, RPA and blockchain technologies. The assimilation of these emerging technologies holds the aptitude of generating smarter, safer, efficient, and more secure systems. Solution providers seem to be more confident in applying several technological developments to answer volatile, uncertain, complex, and ambiguous challenges in the systems [16].

## 1.2 The Problem Statement

In Europe, the initial progress of e-invoicing was driven by the private sector whereas in the present it is being very much supported by the governments [10]. Present e-invoicing business models have been evolved based on traditional paper processing. Therefore, many elements can be enhanced further. Given that the governments are introducing the initiatives for e-invoicing through moves such as mandates [1] and the businesses are requesting the technologies to be upgraded and the services to be better [10], the requirements and opportunities in a comprehensive e-invoicing platform should be first identified.

Even though the e-invoicing in Estonia has reached a considerable level of effectiveness which can also be considered somewhere in between digitisation and automated e-invoicing, there are many drawbacks and opportunities to improve. When considering e-invoicing as a service in Estonia, several barriers such as technological, security-related, organizational and implementational are identified for disruptive innovation concerning real-time economy [12].

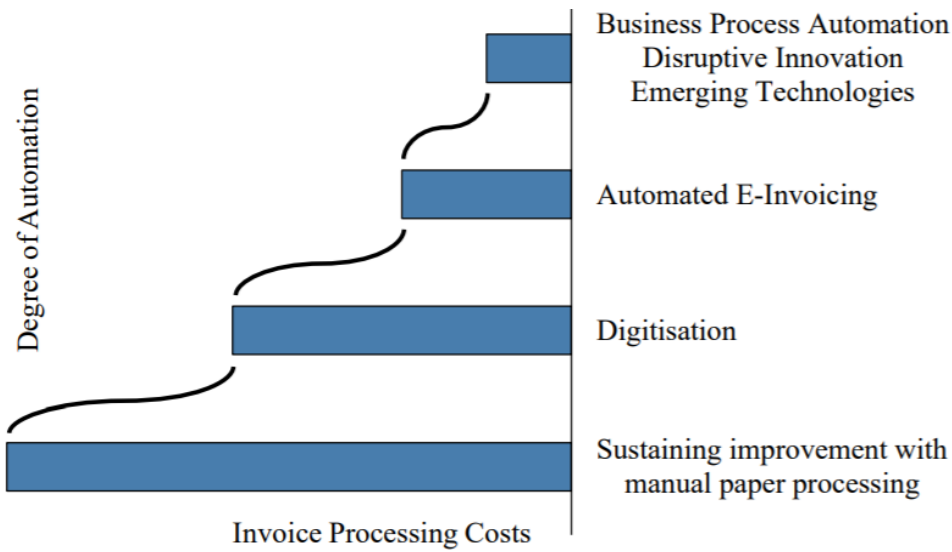

Figure 1 e-Invoicing Development in the European Union [10]

According to Billentis, the above figure shows how automation can be introduced to e-invoicing in a step-by-step manner. And when the degree of automation is higher, the cost of operations is lower. Alternatively, rapidly growing emerging technologies which could be used for disruptive innovation also lay a heavy base for strategic drivers. These technologies take a completely new approach for some of the existing problems and can act as a substitution for old solutions [13],[14]. Where on the other hand one could say that small enhancements in an e-service can only create incremental improvements [11]. Therefore, what is lacking in a mature E-state is a disruptive innovation for a nearly matured e-service such as e-invoicing.

### 1.2.1 Research Question

Against the above backdrop, the key research question emerged as,

*How to enhance e-invoicing systems and increase the efficiency of workflows with higher levels of automation?*

The key research question is divided into smaller questions to derive the hypothesis and make research executable. These research questions aim to find out what kind of inefficiencies are in the e-invoicing systems and process currently, what kind of and to what extent automation is present in the current e-invoicing systems and finally find out how can emerging technologies be applied to eliminate the barriers and achieve higher levels of automation in e-invoicing.

The sub-questions are the following:

RQ1: What are the current inefficiencies of existing e-invoicing systems?

RQ2: What is the current level of automation and what are the requirements for future automation opportunities in e-invoicing systems?

RQ3: What emerging technologies can be used to enhance the e-invoicing solutions eliminating the inefficiencies?

Based on the answers to the previous three research sub-questions, the thesis intends to develop an ecosystem with novel automation ideas for a better e-invoicing system, which will positively benefit all the stakeholders.

## 1.3 Research Aim and Objectives

To find out what are the difficulties or problems in current e-invoicing solutions, understand what methods of automation is present in the current systems and provide insights about how to use emerging technologies as innovation is the main objective of this research.

As another output, the author aims to design a software ecosystem as a novel recommendation to eliminate the inefficiencies and bring in more automation.

A type of stakeholders who will benefit from this study will be the e-invoicing operators, who can enhance their systems with the findings from this study. It is beneficial to understand the current status in the Baltics and build solutions answering these problems. Since the user's perception is included in the study, all the stakeholders who are involved in designing and building e-invoicing and related systems can use the findings. Also, it will benefit the users who are using the systems to understand the problems in them, to engage more in enhancing the systems with the system developers. In general, all the other stakeholders such as government bodies, financial institutes and traders who are related to the above field will be benefitted with the information derived from this research.

According to the recommendations of this study, it will be beneficial to use the designed software ecosystem for creating a prototype of an enhanced e-invoicing system in the Baltics in the future.

## 1.4 Scope of the Thesis

As mentioned in this chapter, the thesis is conducted in the region of Baltics. The experts who are interviewed are from the Baltics region or connected to the regional e-invoicing projects. The users who participated in the survey are the users who use e-invoicing systems in the companies who operate in the Baltic region or in one of the countries in the Baltic region (some companies can be operating at a larger scale than Baltics but this number is very less).

This research is a qualitative study in the scope of e-invoicing systems used in the Baltic region and the methodology is extensively described in Chapter 3.

This research is original and innovative due to several reasons. One major reason is that the author considers two different perspectives to achieve the aims. That is with the professionals who are involved in e-invoicing systems for years and the general users who deal with the e-invoicing systems daily. These two very different populations are exposed to problems in different ways and can have different opinions. The next is that even though there are study conducted on the e-invoicing systems barriers and opportunities, they have not combined it with using emerging technologies to boost effectiveness and bring innovation. The key reason out of all above is the author's contribution to designing a novel and practical software ecosystem out of the research analysis to best of her knowledge on the subject as the recommendation of this study.

The practicality lies in the usability of the study findings in real-life e-invoice system design and implementation. Since the findings have originated from the real users and experts, they are known to be valuable in the subject context. The recommended software eco-system to be feasibility checked as a future suggestion. Even some parts of it can be implemented step-by-step in the future reaching a fully automated system as the final result.

## 1.5 Structure of the Study

The thesis starts with an introduction explaining why the topic was selected and how the research need is identified by the author. The research questions are generated according to the research aim and objectives. The second chapter details the background of e-invoicing. The author presents analysis done on previously published studies concerning each research question which models the data gathering instruments used for the expert interviews and user survey. Chapter three brings out the methodology in-depth and research design. Chapter four is divided into two parts consisting of the thematic study of the expert interviews and the general analysis of the user survey. The analysis is produced inline with the research questions. Chapter five presents the newly designed ecosystem by the author, benefits of the recommendations and the limitations of the research. Finally, in chapter six the conclusion is presented with suggestions for further research.

# 2 Literature Review

The literature review showcases an outline of prior research in the frame of the stated research problem. A critical analysis of the diverse but integrated areas of literature also known as a theoretical content of the study will be brought out in this chapter.

The chapter will give an overview of e-invoicing and its elements. Then it will move on to analysing the literature related to each sub research question. It includes current inefficiencies in e-invoicing systems, current automatization levels in e-invoicing systems and usage of emerging technologies in the realm of e-invoicing and financing. Also, the author has presented published studies on user perceptions of information systems.

## 2.1 e-Invoicing

A commercial invoice is the most significant document exchanged between trading partners. Along with the commercial significance, the invoice is an accounting document which has legal insinuations being the basis of calculating VAT (value-added tax) [28]. On the other hand, e-invoicing is the substitute for traditional invoices in the electronic format, which has even more benefits added to the importance discussed above.

## 2.1.1 E-invoicing Market Maturity

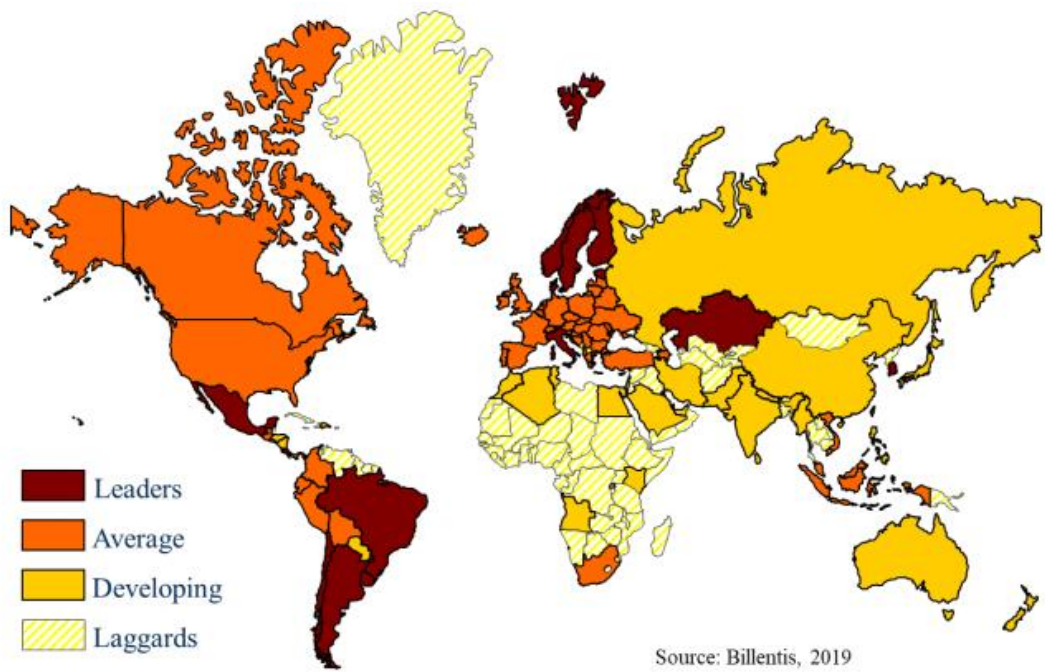

Figure 2 Market Maturity of e-invoices [25]

According to Billentis, the highest market adoption of e-invoices is in Brazil, Chile, Argentina, Mexico, Kazakhstan, Sweden, Norway, Finland, Italy, Estonia etc. When we consider the Baltics, Estonia is one of the leaders in the e-invoicing market. Latvia and Lithuania are in the second-tier with average market penetration [25].

## 2.1.2 E-invoicing Models

There are four main types of e-invoicing models in the e-invoicing landscape according to Koch [10].

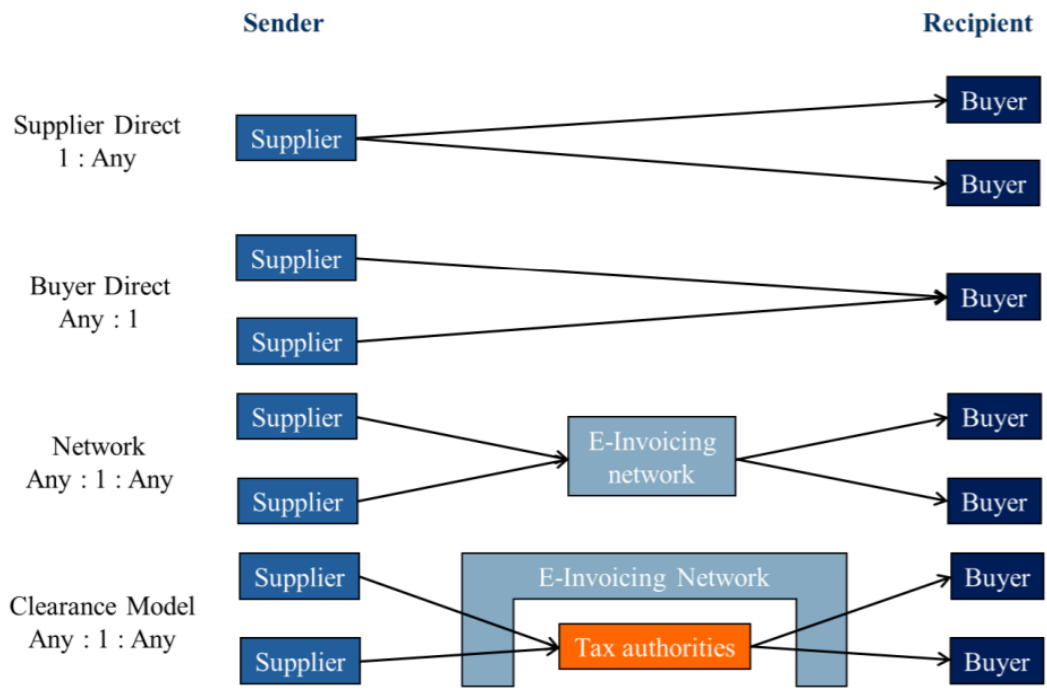

Figure 3 Overview of Main Market Models [10]

- Supplier Direct Model (in-house) - The supplier implements an e-invoicing system in his environment to send out invoices. Customers can log into the supplier's portal and view invoice and download as well. Mainly used by large volume invoice issuers like telecommunication companies.

- Buyer Direct Model (in-house) - The buyer implements an e-invoicing system in his environment to receive e-invoices via different channels. Mainly used by large buyers with limited suppliers.

- Network Model - Buyer and Supplier has an interface to an e-invoicing service provider. Invoice issuer provides data in different formats and the operator translates it to the receiver's format. Operators offer supplementary services, for example, data quality checking and archiving.

- Clearance model - This model has tax authorities connected to buyers and suppliers with or without e-invoicing operators, especially where VAT compliance is required. Some European countries, Latin American countries, and Asian countries are practising this model.

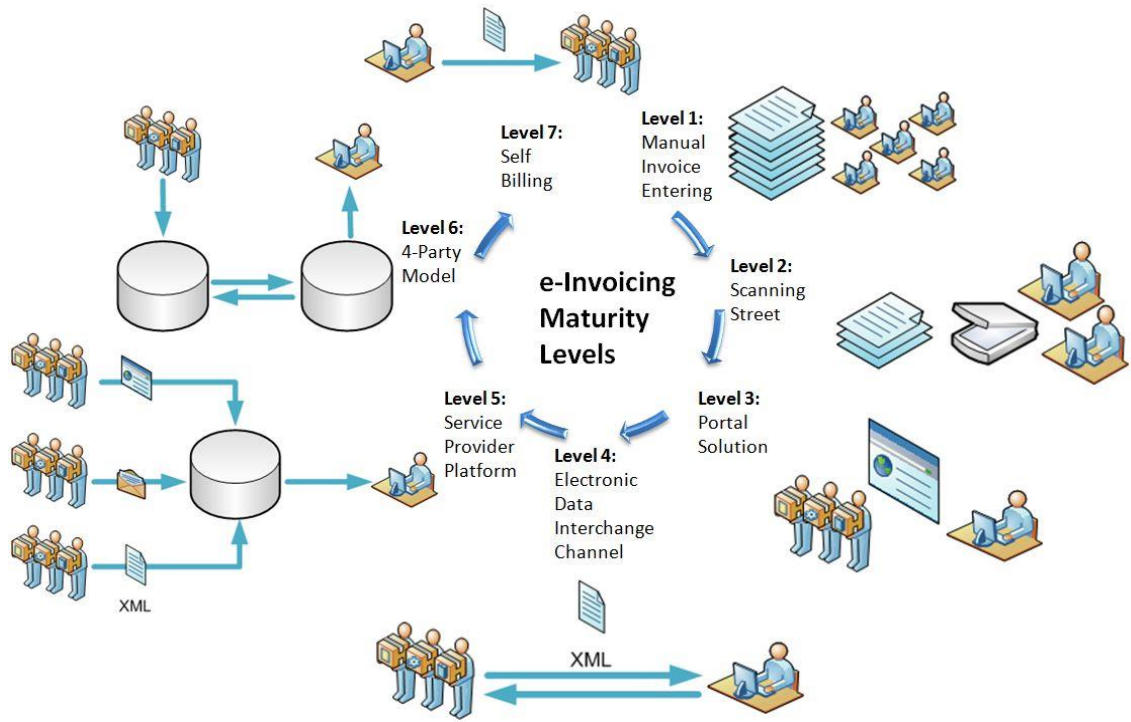

Figure 4 CAPGEMINI Model of e-invoicing Maturity [68]

- Level 1: Manual Invoice Entering

During this level, there are no e-invoicing activities. The invoices are processed manually.

- Level 2: Scanning Street

During this level, the invoice is scanned and digitally available for processing. It still has to be checked manually, yet gives the ability to reproduce the same invoice easily. This also facilitates emailed PDF invoices to be recognized by OCR.

- Level 3: Portal Solution

This is a website where suppliers can log in to see the information about invoices. This level does not require the suppliers to have complex IT solutions. The invoice supplier filled is directly submitted to the buyer's ERP system.

- Level 4: Electronic Data Interchange Channel

This is a huge leap from the previous level. Before it was only EDI connections but now the data is sent by XML file with standards. This channel is good for high volumes of invoices processing. This also brings a lot of cost savings but having a channel for each supplier is paving the way for a complex IT landscape.

- Level 5: Service Provider Platform

The earlier stage difficulties are managed with this level incorporating a special service provider who acts as a hub. Automated checks are performed and translations between standards of messages are done by the service provider itself. If some suppliers or buyers are connected to other service providers should be also connected to their platform somehow.

- Level 6: 4-Party Model

To eliminate the issue of having connected to many platforms, the four-party model came into play. The supplier and the buyer have an e-invoicing service provider respectively. The service providers connect and provide necessary services. This is widely used at present.

- Level 7: Self Billing

This is the ultimate level for a buyer where it is the simplest yet effective method. The buyer makes agreements with the supplier, once the goods are received the buyer itself can create the invoice and do the payment. This model cannot be implemented at all occurrences, additional rules and regulations must be present as well.

[68]

Currently, in the Baltics, the 4-party model and minor levels from that are being used evidently.

### 2.1.3 Types of Digital Invoices

An electronic invoice is a form of a digital document, where digital is described as "composed of data in the form of especially binary digits" [17].

In many instances, there is a confusion between electronic invoices and other phenomena related to the digitization of invoices. Described below are a few categories of invoicing and in this study, the author concerns about the fully electronic invoices.

Table 1 Categories of Invoices [18]

| S. Sales Invoices | P. Purchase Invoices |
| --- | --- |
| Fully Electronic Invoices | |
| Invoices sent directly through web e.g. e-invoice to net bank | Invoices received directly through web e.g. e-invoice to net bank |
| Invoices sent directly to electronic purchase invoice processing system e.g. EDI invoice | Invoices received directly to electronic purchase invoice processing system e.g. EDI invoice |
| Email Invoices | |
| Invoices sent by email or as an attachment of the email directly to the recipient. | Invoices received via email or as an email attachment. |
| Hybrid Electronic and Paper Invoice Combinations | |
| Invoices are written with an in-house computer but are in a PDF for a similar format. | Invoices received on paper but scanned for electronic processing |
| Pure Paper Invoices | |
| Invoices are written or printed and enveloped | Invoices received on paper and processed as paper invoices |

### 2.1.4 e-Invoicing Standards

Directive 2014/55/EU [26] for Electronic invoicing in public procurement is the main guidelines for European member states. This directive calls for a definition for a common standard in Europe on e-invoicing semantics. Also extra standardisation on interoperability with the syntaxes. Since all the Baltic countries are member-states of the European Union, they must follow the above directive.

According to Billentis [25], the deficit on knowledge about existing standards and pridefulness of reclusive establishments are the reasons that stakeholders not using the standards. And this has paved the way to re-inventing different niche standards domestically.

The table underneath describes few standards used in Europe and elsewhere.

Table 2 International and Industry Independent Standards for Electronic Business Messages [25]

| Name of the Standard | Explanation of the Standard |
|---|---|
| ebXML | ebXML (Electronic Business XML) is a project to use the Extensible Mark-up Language (XML) to regulate the exchange of business data securely. Also, ebXML assist companies to manage business relations, corresponding information in mutual standings, define business processes.[19] |
| Oasis UBL 2.x ISO/IEC 19845;2015 | "UBL, the Universal Business Language, defines a royalty-free library of standard XML business documents supporting the digitization of the commercial and logistical processes for domestic and international supply chains such as procurement, purchasing, transport, logistics, intermodal freight management, and other supply chain management functions" [20]<br><br>UBL is used by the standards for "PEPPOL (Pan-European Public Procurement On-Line) platform" [20] and many public procurement initiatives. |
| UN/CEFACT | UN/CEFACT is a body of the United Nations, which is the answer for "lack of an international data exchange standard for invoicing"[21]. "UN/CEFACT addressed this critical need with the release of the Cross-Industry electronic Invoice (CII).4 The CII can be used, for instance, by the Steel, the Automotive or Electronic industries, the retail sector or Customs and other Government Authorities." [21].<br><br>This standard has established,<br>- "The UN layout key for trade documents (EU's Single Administrative Document – SAD)<br>- UN/EDIFACT the international standard for electronic data interchange<br>- UN/CEFACT XML<br>- Trade facilitation recommendations" [25] |

| Name of the Standard | Explanation of the Standard |
|---|---|
| PDF/A-3 ISO 19005-3 | PDF/A is an iso standard for PDF particular for "for preserving the static visual representation of page-based electronic documents over time in addition to allowing any type of other content to be included as an embedded file or attachment." [22]. |
| European norm 16931 CEN/TC 434 | The Directive 2014/55/EU needed to develop a common standard for EU, for e-invoicing in B2G buying to remove the cross-border blockades. This consisted of "the European standard on the semantic data model for the core elements of an electronic invoice, a technical specification on a limited number of invoice syntaxes and other components. Two syntax formats approved by CEN are UBL and UN/CEFACT" [25] This is the standard which is supported by all EU public entities. The speciality of this is it supports country specifications eliminating some elements. Besides, it supports country or businesses specific extensions uplifting the B2B sector [23]. |
| CEN/TC 440 [24] | The main aim of this standard is "to support and facilitate the electronic public procurement processes and their underlying accompanying information flows in the physical and financial supply chain. It considers standard messages for e-notification, e-tendering, e-ordering and e-fulfilment". [25]. |

The integration of many e-invoicing standards is generally out of the scope of ERP and accounting software providers. Hence it has become the main responsibility of third-party e-invoicing service providers to format the data differently to suit each standard, which makes life very hard for the service providers [25].

## 2.2 Research Model Analysis

The research model consists of the elements which will measure the sub research questions. To derive the elements (key elements) for each sub research question, published literature will be analysed.

### 2.2.1 Identifying Current Inefficiencies in a Solution

To assess the effectiveness of an e-invoice system many methods of measurement are possible. The same can be applied to find out the inefficiencies of a current system.

To understand the above the indicators are taken in the means of functional and non-functional which is one major classification in software systems [27].

According to the European Commission study; functionality related to e-invoicing is described under topics such as "automatic processing of e-Invoices by the system, invoicing specific functionality, archiving specific functionality, data requirements, dispute specific functionality, functionality regarding the querying and retrieval of information" [37]. Each of the topics is further detailed into the user story level which the author of the study thinks has quite a good coverage of functionality. According to Kiroski, Gusev and Kostoska; the functionalities of e-invoicing with e-delivery are identified as user administration, export capability, search facility, report facility, taxations, document types and delivery options and item details[36] which the author of this study thinks that it does not cover all the functional requirements compared to the European Commission study. A study done by PricewaterhouseCoopers points out functional requirements such as retention of the e-invoices, identification of the partners, e-invoicing process requirements and legal requirements [38]. According to Radeski and Davcev, the functional requirements are specified as required XML elements and format, electronic signature enabled documents, delivery notifications, automatic transferring and receiving of data, dispute mechanisms, user administration and tax audit controls [39]. This study also covers the major functional requirements of e-invoicing but the concerns like digital archiving and cross border compatibility are not discussed.

When it comes to non-functional elements, a study by Karantjias, Papastergiou and Polemi points out elements such as interoperability, security and trust, scalability and extensibility are the non-functional elements that an e-invoicing system should have [40]. Kiroski, Gusev and Ristov discuss the non-functional elements such as elasticity and scalability, simplicity, cost-effectiveness, business agility and resiliency concerning an e-invoicing IaaS (Infrastructure as service) cloud model [41]. When e-invoice systems are being served by service-oriented architecture, the following non-functional elements can be discussed such as interoperability, performance, security, availability, modifiability, usability, scalability and reliability [42]. According to Kiroski, Gusev and Kostoska; the non-functional elements for an e-invoicing system emerge as security, technical capabilities, transparency, adaptability [36]. According to Delone and McLean, the system quality is measured with elements adaptability, availability, reliability, response time and usability in their Information Systems Success Model 10 years update [43]. In Taiwan's e-invoicing system Joung et al. describe security and privacy concerning the non-functional elements of the system [44].

**2.2.2 Identifying Current Automation Level in a Solution**

About automation in the current e-invoicing systems, it is evident that the automation is present at some level according to Ingrid Pappel et al. One such example is the system automatically capturing some invoice fields and registering into ERP [45]. In par with the above example, Pappel and Kosenkov have described the application of e-invoicing in industry 4.0 where industry 4.0 is a concept highly cohesive with automation. They have figured out four types of barriers for implementing industry 4.0 such as technological, security-related, organizational and implementational [46]. In the Billentis report, it is discussed that "one of the main challenges of e-invoicing is business process automation"[47]. Also, it is mentioned that "automation paves the way for real-time data validation" [10]. According to Caluwaerts, the degree of automation present in a system directly affects the cost-saving [49]. Igor Pihir and Neven Vrček have analysed requirements for a fully automated e-invoicing system for SME companies, in which they have paid attention to how an e-invoice acts concerning different parties such as trading parties, customs and tax authorities [50]. According to Gibser et al, the industry 4.0 and automation related to that will predict to be resulting in a digital ecosystem which has even more integrated and flexible value chain networks. The automation in a system can be identified with technological effectiveness and functional effectiveness [48].

**2.2.3 Emerging Technologies: Blockchain**

Blockchain technology is one of the latest emerging technologies in modern economic solutions. "The blockchain is an incorruptible digital ledger of economic transactions that can be programmed to record not just financial transactions but virtually everything of value" is the definition of the technology constructed by Don and Alex Tapscott [29].

The main and the most important advantage of blockchain is being decentralised, which eliminates the need for a central administrator or a third party to manage and all the parties related can make informed decisions with trust [30]. The blockchain refers to a "public accessible distributed ledger that ensures the integrity of all kinds of transactions" [35].

As an alternative to a single official ledger that stands for a particular transaction took place, this common ledger can reproduce the information it carries on several nodes. When the number of widely held ledgers are more than contaminated or tampered ledgers, the trust would be on these multiple copies. Therefore, a centralized administration for authentication is not needed anymore. Even though the data is decentralized establishes a feasible option the overriding entities should be managed as to identify which data should be lawfully stored. When we explain this from a technological angle, the selection of which universal truth is selected to keep posted on the whole blockchain network should be well-defined [31]. "Whereas replication is the key to decentralization, a so-called distributed ledger represents the concept to provide the functional infrastructure among equipotent participants" [35].

The blockchain is the initial useful application to create a completely "public permissionless distributed ledger"[35]. The capability of creating trust among different stakeholders are connected to the immutability quality and it further connects to how information is organised, created and dispersed. Consequently, the blockchain technology amalgamates years of study and is constructed on four key concepts [35],[32].

1. Peer-to-peer network: the database structure for the blockchain which is decentralised with this network architecture.

2. Transaction logic: encryption and a digital signature are used to protect the transaction process between anonymous accounts.

3. Data immutability: the ledger contains data blocks which are cryptographically secured and linked to preceding data

4. Consensus mechanism: an algorithm allows for a universal selection, permitting the users to reach agreement on a true systemic state of the blockchain to synchronize the distributed ledger.

Concerning the published studies on using blockchain in e-invoicing, these studies were analysed by the author of the thesis.

Table 3 Published Studies on Blockchain

| Authors | Year | Title | Description |
| --- | --- | --- | --- |
| E Hofmann, UM Strewe, N Bosia | 2017 | "Supply chain finance and blockchain technology: the case of reverse securitisation"[51] | Describes BCT being incorporated in Supply Chain Finance. Investigates what are the possible opportunities in approved payables financing. The study has identified possible pain points and barriers. Then a blockchain-driven model is derived where the uses and opportunities of the technology are discussed. Further, it discusses the tokenizing of invoices to reduce duplicates and frauds across networks [51]. |
| A Faccia, MYK Al Naqbi, SA Lootah | 2019 | "Integrated Cloud Financial Accounting Cycle: How Artificial Intelligence, Blockchain, and XBRL will Change the Accounting, Fiscal and Auditing Practices"[86] | BCT can assist e-invoicing by eliminating the need for recognition of graphic signatures, allowing the subscriber to be identified uniquely [86]. |
| L Segers | 2019 | "The design of an open, secure and scalable blockchain-based architecture to exchange trade documents in trade lanes"[52] | How smart contracts are implemented, elimination of manual cross-validations. Especially focusing on how to aggregate information from different sources and automatically generate a customs declaration. The study finds that "an open, secure and scalable blockchain-based architecture can support the exchange of trade documents" [52]where some of these documents are invoices. |

| Authors | Year | Title | Description |
|---------|------|-------|-------------|
| Lawlor C | 2016 | "Tokenization of invoices: a blockchain technology supply chain finance use-case" [53] | Each invoice should be distributed such as a bitcoin transaction hashed and time-stamped, which creates a unique identifier. Double financing would be eliminated with this practice [53]. |

**2.2.4 Emerging Technologies: Robotic Process Automation (RPA)**

The Robotic Process Automation (RPA) is a novel dimension of upcoming technologies. RPA can be also introduced as cutting-edge technology in the realms of computers science. According to Madakam et al, "it is a combination of both hardware and software, networking and automation for doing things very simple"[33].

The Vice President of Research at the Everest Group Sarah Burnett has stated that "Robotic Process Automation is the next wave of innovation, which will change outsourcing. We already are seeing the beginnings of a race to become the top automation-enabled service provider in the industry. In time, we are likely to see an arms-race for innovation in automation tools leading to new offerings and delivery models."[33]

There are following examples of usage of RPA.

- Repetitive processing of transactions (e-invoice processing as an example)
- Rules-based processing (error checking as an example)
- Handling a large number of transaction volumes (orders processing as an example)

RPA brings direct cost-effectiveness while enhancing precision across establishments and businesses. Software robots who are trained to perform a variety of repeating functionalities can interpret, prompt replies and connect with different interfaces or systems just like human beings. Software robots are known to be "only substantially better: a robot never sleeps, makes zero mistakes and costs a lot less than an employee" [33]

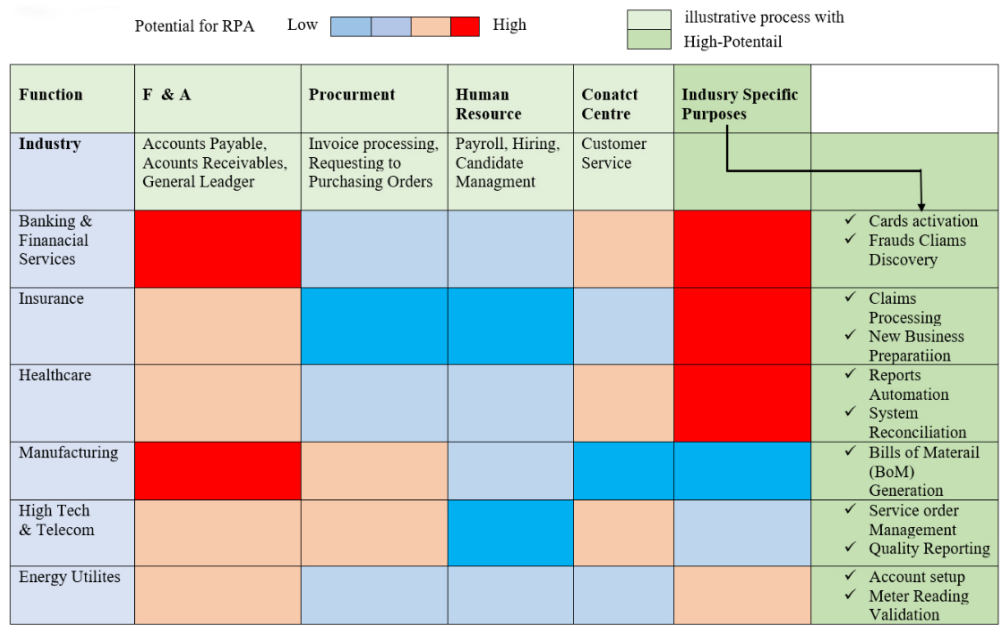

Figure 5 RPA Adoption by Industry and Function [33]

According to the above figure, the probability of incorporating RPA in e-invoicing is very high [33].

It was predicted that the market value of RPA technology and applications are up to $400,000,000 by 2017 and it is anticipated to rise to nearly $1200,000,000 by the year 2021. The projects of RPA consist of implementing RPA and consultations which focus on constructing RPA competences inside a company. RPA is made out of bots and algorithms. Automation and AI are aimed at reshaping the finance function in the future. This tells that there will be many automated-devices fixed on top of financial institutions or banks to carry out day to day repetitive tasks to reduce the operational cost of everyday online-services and handling security. These bots are known to be extremely user-friendly who make sure that the operations are accurate. One example of using RPA in the financial sector is the India State Bank of India. Gartner predicts that by the year 2020, customers are to manage 85% of their associations with the companies deprived of cooperating with a human being. Which concludes that the future holds a lot of automation [33].

Concerning the published studies on using RPA in e-invoicing, these studies were analysed by the author of the thesis.

Table 4 Published Studies on RPA

| Authors | Year | Title | Description |
|---|---|---|---|
| Muurinen Juho | 2019 | "Robotic Process Automation as an Automation tool for improving Purchasing processes"[54] | "This is empirical research on how RPA is adopted by a company for supply management and operative purchasing. The study highlights several benefits and challenges faced in the RPA projects. The study also presents factors related to choosing an automated system for the business process" [54]. |
| Vipin K, Suri Marianne, Elia Jos van Hillegersberg | 2017 | "Software Bots - The Next Frontier for Shared Services and Functional Excellence"[55] | "The study has analysed where RPA is deployed at shared service organizations and described the business case. The drivers and challenges are being also identified in the research. The outcome is the functional processes where RPA can be applied and the implementation steps" [55]. |
| SARA LAAHANEN, EEMIL YRJÄNÄ1 | 2019 | "FINTECHS: THEIR VALUE PROMISES AND DISRUPTIVE POTENTIAL"[56] | "The study brings out real-life applications of Fintech technologies. How robot advisory is performed with mathematical rules and algorithms which are used for investment advisory services and asset management" [56]. |
| Lintukangas, Annika | 2017 | "Improving the indirect procurement process by utilizing robotic process automation"[34] | "The research analyses how RPA is utilized in the indirect procurement process. It investigates which tasks robotic process automation can be utilized and what are the benefits and disadvantages of the usage" [34]. |
| Carlo Aalberto Visentini | 2019 | "The future of procurement: adoption of robotic process automation and responsible purchasing practices." | "Is a case study on how companies such as Unilever, Vodafone and Adecco has automated their procurement processes with RPA" [58]. |

### 2.2.5 Emerging Technologies: Artificial Intelligence

Artificial intelligence (AI) has been studied extensively from that time the Turing Test was created in 1950. There were periods where the interest in AI was low (known as AI winters) and conversely there were periods where AI was on-demand gaining rapid advancement. From the 2000s, with the increasing computing power combined with the introduction of machine learning and big data, the realm of AI witnessed success stories such as IBM Watson and Deepmind's AlphaGo2 [59].

In 2016, at the World Economic Forum, the significance of AI was stressed and the impact of AI on the governments where it makes the governments agile [60].

AI is used in non-conventional areas such as human resources and recruitments and also conventional areas but has not been enhanced such as in financial predictions. There are different methodologies and techniques of AI which can be classified as,

1. Techniques which reply on the mathematical enhancement
2. Network-based approaches wherein represents problems are sets of possible states and transitions in between them
3. Agent-based methodologies and multi-agent system interactions
4. Automated reasoning-based approaches on present knowledge
5. Big data and machine learning analytics

[59]

Concerning the published studies on using AI in e-invoicing, these studies were analysed by the author of the thesis.

Table 5 Published Studies on Artificial Intelligence

| Authors | Year | Title | Description |
|---|---|---|---|
| SARA LAAHANEN, EEMIL YRJÄNÄ1 | 2019 | "FINTECHS: THEIR VALUE PROMISES AND DISRUPTIVE | "How predictive analysis and AI can be used for things such as fraud detection, investment decisions and making client risk profiles" [56]. |

| Authors | Year | Title | Description |
|---|---|---|---|
| | | POTENTIAL"[56] | |
| Stefanovova, Z., Bartkova, H. and Peterkova, J., | 2020 | "Evaluation of the Effects of Digitization in the Process of Accounting Operations in a Selected Manufacturing Company"[57] | "This study shows how AI can assist processes in e-invoicing such as read invoice automatically and record it in the accounts. It also analyses the problems regarding the above" [57]. |
| Violeta SIMIONESCU | 2016 | "The Impact of Artificial and Cognitive Intelligence on Romanian Public Procurement" [60] | "The paper analyses the AI and cognitive intelligence trends concerning public procurement. It also discusses a social system for public procurement than a mechanistic system in the context of Romania" [60] |
| Nigel Cory | 2020 | "Why Countries Should Build an Interoperable Electronic Invoicing System into WTO E-Commerce Negotiations" [61] | "In the context of e-invoicing, one point this paper ponders upon is that governments can fight tax fraud with AI cross-checking a vast number of transactions" [61]. |
| Guy Berg | 2020 | "Electronic invoicing: The first step towards digital B2B payment modernisation" [63] | "This research described how to modernize the payment systems in the US regarding end-to-end automated processes for electronic payments. Sending structured electronic invoices that can be automatically processed by the buyer, paying the supplier electronically, sending sufficient payment remittance information electronically to link the payment back to the original invoice and providing better connectivity and interoperability between businesses to enable the previous three factors are the key findings" [63]. |
| Nistoreanu Puiu, Ene Irina | 2019 | "Consumer Perception Regarding the Role Of AI: A Discriminant Analysis Based on Age" [62] | "The study analyses the consumer perception towards AI. It is an age-based analysis where people below 40 are open towards new technologies and trust is high on the use of AI. People above 40 did not have trust in using AI-based systems and not open for new technologies."[62] |

## 2.2.6 User Perceptions Towards Using an e-invoicing System

In this subchapter, the literature regarding user perceptions towards e-services and e-invoicing systems are discussed. Jiunn-Woei Lian has described seven "Critical factors for e-invoice service adoption in Taiwan according to Unified Theory of Acceptance and Use of Technology 2(UTAUT2)" [69]. It was decided that five of those factors to be used to measuring the user perception in this thesis. Out of the 7 factors, social influence and facilitation conditions are taken out assuming the users work in a corporate structure where all the facilitation conditions are fulfilled such as internet connections, training on the system etc and there is no social influence in a corporate environment, as it is a business requirement to use an e-invoicing system at the workplace.

Perceived Ease of Use/ Effort Expectation

Perceived ease of use is "The degree of ease of use with the use of e-invoice" according to Davis [70], Venkatesh et al [71] and Walczuch et al [72]. Even it can be defined as 'the degree which the user can effortlessly use an e-invoicing system' where it has an inverse relation with the notion of complexity [70].

Perceived Usefulness/ Performance expectation

Perceived usefulness is a key variable in most of the TAM (Technology Acceptance Model) and its derivatives. It describes the degree to which users of e-invoicing systems will consider that its use will improve their performance, also providing advantages in the organizational level [70].

The attraction of e-invoicing systems is the advantages of these systems. It can determine the usefulness perceived by the users and the firms. Perceived usefulness can modify user behaviour and predict the continuous usage of the service [73].

Perceived Risk

Perceived risk is known to be "the degree to which people perceived risk when using e-invoice" [76].

Perceived risk is known to be affected by the perceived trust. Then the perception of risk is affecting the actual usage of a technology or a system [70]. There are many different types of perceived risks such as functional, physical, financial, psychological and social.

Perceived risk can involve possible social consequences, financial loss, physical danger, loss of time and ineffective performance [77].

Perceived Trust

In the context of this study, the trust is "the degree to which people trust government e-invoice policy and service" [74]. Some researchers have analysed the role of trust in different stages such as trust before the use of the system (pre-use trust) and after use of the system (post-use trust). In the pre-use the users are not familiar with the technology and the systems, hence trust is based on their tendency to trust. Once the users have used the system or the software, experience and previous perceptions determine the level of trust they have [75]. The thesis analyses post-use trust since the users is already using some sort of e-invoicing system.

Perceived Security

Deficiencies in security are one of the most noteworthy barriers in the development of e-commerce related software products. If the knowledge of IT systems is low in users, the fears of hackers or a third party will tamper the information is high. Security reflects a perception of reliability of the methods of data transmission, storage and access [69].

E-invoicing must adhere to the requirements imposed by Directive 2011/115/EC. These requirements include the relationships with other agents in terms of authentication and non-repudiation of transactions. Other requirements of security can be derived from company policies and national level requirements such as archiving digital content for a while [28].

# 3 Research Methodology

This chapter focuses on bringing a comprehensible research design into place. It talks about the incorporated theoretical standpoint regarding the research combining the formation of research methodology and associated strategy.

The chapter elaborates the rationale behind the research methodology, the confirmed research approach and data collection techniques incorporated to find a solution for the stated research question. The analysis techniques of data will be defined which eventually leads to data interpretation in the following chapter.

This chapter will explain in detail how the research was conducted. First, the choice of qualitative research will be discussed. Subsequently, it will be explained how the data was gathered and, lastly, how it was analysed.

The research process is a methodical way of collating and interpreting data gathered to create new knowledge in an investigative way of presenting reliable discoveries [64]. According to Hair, there are three phases in research, namely; formulation, execution and analytical. During the formulation phase, the need for the research is confirmed and the problem is defined. The execution phase consists of research functionalities to gather data defining the sampling methods, data collection methods, collecting data, coding data and storing data. The third phase contributes to the analysis of data gathered in the execution phase, interpreting and performing inferences in line with the research limitations. At the end of the analytical phase, the academic contribution is made which finally establishes the new knowledge [65].

## 3.1 Research Philosophy

A research study design should begin with a selection of a topic and a paradigm. A paradigm in a holistic view of beliefs, values and methods within the scope of research [66].

"A qualitative study can be defined as an inquiry process of understanding a social or human problem, based on building a complex, holistic picture, formed with words, reporting detailed views of informants, and conducted in a natural setting" according to Cresswell [80].

Qualitative research is known as an inductive process of data arrangement into different categories and seeing relationships between the categories. This means that data and their meaning are gathered originally from the research context. The interaction between variables is evident in qualitative research. Open-ended questions are used to derive detailed data in the interviews and structured questions are used to mine the user perceptions through a user survey. Qualitative research is a broad term for different approaches and methods in doing research such as ethnographic, naturalistic, anthropological, field or participant observer research [81].

### 3.1.1 The Rationale for the Methodology

As stated in Chapter 1, with the increase in demand on invoicing operations, it is essential and inevitable for Baltic regional countries to adapt to electronic invoicing applications to supply the demanded coverage. As referred by the author, even though the e-invoicing in Estonia has reached a considerable level of effectiveness which can also be considered somewhere in between digitisation and automated e-invoicing, still there can be multiple loopholes and bottlenecks among the current e-Invoicing applications used within the Baltic region which can be improved and enhanced. Certain barriers and drawbacks such as technological, security-related, organizational, operational etc will be researched and discussed under the thesis objectives.

Qualitative data analysis frameworks will be incorporated to study the research data as the research will require subject matter expert opinions and feedback via coordinated interviews and perceptual feedback from general e-Invoicing application users via publicly shared surveys.

Upon data gathering, the author will analyse the data via thematic analysis and observe if the research questions mentioned in Chapter 1 are answered or not. (Will be explained extensively under 3.2.1 Research Approach section)

## 3.2 Research Design

Research is carried out according to a design which is the main tool of a researcher. The researcher must strategize the procedure of the research which would be steered by the philosophy, research methodology concerning the literature and access to data [65].

**3.2.1 Research Approach**

To answer the research questions, the author selects an inductive approach as a research approaching method.

Initial observations will lead to developing conclusions related to sub-research questions and to develop an enhanced e-invoicing ecosystem with emerging technologies incorporated as a suggestion. There will not be any theories or hypotheses applied under this approach [78]. Upon pattern identification, the author will develop empirical generalizations and will tend to identify primary relationships among the patterns [79].

Under an inductive approach, a triangulation method is used to validate the research findings. Triangulation helps to ensure that the fundamental biases generated from a single method or via a single observer upon combining theories and methods are overcome [79].

From the generalized triangulation methods, methodological triangulation which emphasizes the usage of multiple data collection procedures such as interviews and surveys will be incorporated in this research study.

In pursuance of identifying interview data, a systematic approach under thematic analysis is followed. Under this analysis, patterns among the interview data are explored iteratively. There are a few steps to conduct a thematic analysis [78],[67].

- Getting familiarized with data - This is the initial phase of the analysis where the author will have to deal with data gathering from conducted interviews and then transcribing them into text format for analysing purpose.

- Establishing codes - Once the data is set, the author can implement codes on transcribed data to identify anything interesting among the feedback given by the interviewee for data analysis. In this way, it will be much easier for the author to organize data in a structured manner

- Establishing themes - Themes will be the next level of codes where the author collates codes into much broader themes for analysing purpose. During this process, some identified codes might be removed or added

- Reviewing themes - Themes refinement happens in this phase as certain themes will need to be split into multiple sub-themes for better analysis. This process will need to be continuously executed until the refined themes become coherent and distinctive.

- Defining and naming themes - Once the themes are refined, the author names the themes in a manner where the theme will be descriptive enough to identify the relationship between the rest of the themes to build a theory.

- Report production - Based on the result from the data analysis, conclusion for research questions can be deducted under this step. Themes will be soundly explained to elaborate on the research findings and to come up with the conclusion.

## 3.3 Population

There are two populations used for this thesis. Both populations have a binding trait being exposed to e-invoicing systems.

The research population to analyse user perception are the users of e-invoicing systems in the Baltic Region. The research population for analysis of inefficiencies in the technology used for e-invoicing systems and the possibilities of using emerging technologies in the field of e-invoicing are the experts on e-invoicing (also in the Baltic region) who expressed their opinions.

For the sampling method, a non- probability sampling method has been used for the survey. The sampling method is a mix of convenience sampling (sample includes individuals who happen to be most accessible to the researcher) and voluntary sampling (sending a survey via e-mail and posting on related LinkedIn forums).

For Experts interviews, purposive sampling has been used where the researcher used her judgement to select most useful to the aims of the research.

*Check Appendix 1 for Interviewees Description.*

## 3.4 Data Collection Methods

### 3.4.1 Interview Questionnaire Outline

Interview questionnaire is based on the three sub-research questions of the study. For each research question key elements are derived from the literature review under chapter 2.4. And the interview question is built on the key elements. This was done to gain depth for each question and to cover the research question as a whole.

*Check Appendix 3 for Interview Questionnaire Outline.*

### 3.4.2 Survey Questionnaire Outline

The survey questionnaire consisted of two main parts. The first part was to understand the demographics of the users. The second part was to understand the user perceptions towards e-invoicing and to finally understand as a customer the level of automation expected in e-invoicing and specific automated features users are willing to use.

## 3.5 Data Analysis Tools

Data analysis is performed with Nvivo 12 software for qualitative analysis.

For the survey results, excel sheets have been used to analyse data and draw the charts. For each country's analysis of the survey, the results of each country are calculated separately with simple excel formulas.

# 4 Data Analysis and Discussion of Findings

## 4.1 Analysis of the Interviews with Experts in e-invoicing

The thematic analysis was done as an inductive method, therefore there is less opportunity to miss some valuable concepts brought out by the interviewees. Also, the full thematic map is presented in *Appendix 2*. For the ease of explaining and understanding, the thematic map is divided into three parts according to the sun-research questions.

### 4.1.1 What are the current inefficiencies of existing e-invoicing systems? (RQ1)

The themes emerged for RQ1 with several sub-themes listed below

1. Information Security (referred to as InfoSec)

2. Technological

3. Operational

Table 6 Thematic Analysis Nodes and References for RQ1

| Sub-Theme | Number of Nodes | Frequency of References |
| --- | --- | --- |
| Information Security | 6 | 13 |
| Technological | 3 | 10 |
| Operational | 13 | 58 |

Throughout six interviews, the number of nodes for each sub-theme and the frequency of references to each category (node) of problems are depicted by the above table. Most of the e-invoicing problems can be categorized as operational problems.

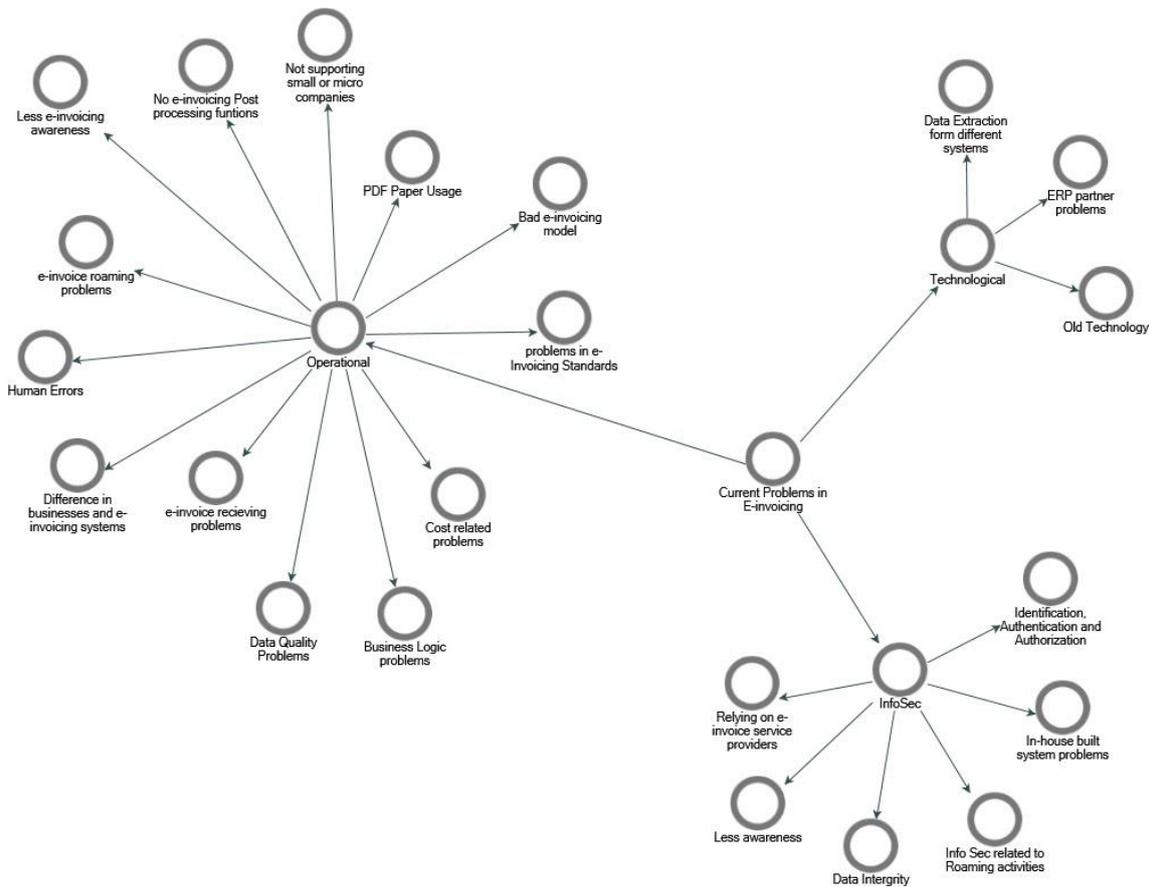

Figure 6 Current Inefficiencies in e-invoicing in the Baltics (Thematic Map)

Information Security (InfoSec)

When we take the theme 'InfoSec' which stands for 'Information Security' related inefficiencies, problems related to several codes emerged. There are several authentication, authorization and identification problems discussed by the interviewees. It is evident that there are problems such as no electronic signing, not having fully automated systems, some still arrive as PDF or Paper and getting extracted to e-invoicing systems and having to manage different channels of access to a system.

Also, it was mentioned that there are data integrity problems, that the information can be changed during the transportation layer. Currently, companies just rely on e-invoicing service providers for data integrity. There is another major functionality that the e-invoice operators perform which is e-invoice roaming. This is not controlled by any authority currently. An interviewee mentioned, *"But there are only few e-invoicing service providers in the Estonian market and we know each other very well and the security is handled by each party"*. The problem arises when the market grows or broadens outside Estonia or Baltic region respectively.

Regarding the small systems built in-house by the companies, they lack information security aspects. One of the interviewees gave an example *"I asked a question at a conference. How many participants are doing penetration tests for e-invoicing systems and there were only 2-3 hands from about 100 participants at a conference. Normally people do not do security-related tests for in-house built systems."*

Technological

One of the main problems identified is the technologies and systems used for e-invoicing in the Baltic region are pretty outdated. Different kinds of examples for the above were given by the stakeholders. *"The biggest problem in every company is legacy software. Even our company has some legacy software and, in some cases, the logic is not correct anymore. We are developing a part of our system and it is very difficult and takes a lot of time because the new system should work with the old system and there are a lot of exceptions. In some cases, it is very hard to upgrade even."* And the technology in the transportation layer seems to be old. *"I think it's not old but the open issue is there is old transportation. Example Fitek is using quite a lot of old SOAP connections."* Another classic example was *"For example, one company in Lithuania still uses Windows XP and we cannot as an e-invoice provider cannot upgrade our REST API channels there."* for clients using old software.

The next significant issue in technology is the ERP partners who are connected to e-invoicing systems. In the e-invoicing realm, the accounting system or ERP system should be connected to the e-invoicing system to achieve automation and other benefits. Some ERPs are quite costly to be connected to an e-invoicing service provider. One example brought out by an expert was *"If it's some global product such as Microsoft product (NAV), then in some cases the ERP connection to service provider interface can cost you some 5000 or 10000 Euros and for a medium-sized company it can be too expensive to connect to the network."*. ERP systems can be less capable when it comes to integration and fulfilling the requirements of e-invoicing. Some findings are *"The problem is MERIT ERP is sending out Estonian e-invoicing which is not enough for EDI in many cases. Merit said they have no more information in the Merit system. Simply if there is a big* *company like Maxima on one side with huge but well-defined business processes and the other party is a small company who is using Merit there will be a problem."*, *"Another problem is there are so many (in Estonia more than 100) different accounting and ERP systems being used, most of them do not have sufficient checking regarding the information added to e-invoices. In the end, we have different quality of invoices."*

E-invoicing systems still use data entering manually or extracting with digitization services which are not that accurate as stated by the experts. It is mentioned that there are no index files when the information comes through PDF or other formats.

Operational

Operational inefficiencies are the most category out of all the inefficiencies. One massive problem is with the e-invoicing standards where all the interviewees mentioned it. It is mentioned that the standards are not strongly imposed as they are in payments, securities trading or telecom. Some companies are still using older standards and there are compatibility issues with the formats and the data fields. One expert brought the issue out as, *"Especially as we take this, Pan-European Public Procurement Online or PEPPOL. In PEPPOL you have very strict validation rules. You take this field and it defines everything in the file. Some address type coding. Is it some VIP rule? Is it some checking cross amounts? and so on. But in Estonian standard, we have just basically data validation rules, which are XSD files. So, yes, there is some date and amount. But is it the correct amount? Is it logical compared to others? So that's the problem."*. Also, some of the versions of standards are old and companies are not using the latest version of a standard. Another standard related issue is that all of the standards do not cater to every type of businesses. An example of this would be *"When we take Volkswagen for an example in the German car industry requiring e-invoices for car parts, the European e-invoicing does not cover all the requirements needed. This means they are still using EDIFACT because they cannot use the European E-invoicing Standard."* Also, standards can be loosely validated, the receivers might end up interpreting non-mandatory field data in their way, paving the way for larger problems.

Even though the study is on e-invoicing systems, a lot of instances are incorporating PDF or images as invoices that cost a lot to be digitized. They can be erroneous, not validated and inefficient. A lot of PDF and other image files are circulated in Latvia and Lithuania than Estonia.

One expert stated that e-invoicing operators are only supporting mid and large-sized companies but not small or Micro companies.

Several functional elements are not supported for the e-invoicing process and one of them is not having an approval process in the system after receiving the e-invoice. In Lithuania, this problem is seen according to an expert *"Lithuania government did their public self-service platform 'eSaskaita' and told to send e-invoice via the portal. There are many drawbacks in this, that there are no approval flows afterwards they print and do the approvals manually"*

Another problem under the theme operational is that the stakeholders have less awareness about e-invoicing. The information contained in an e-invoice and the structure is not known by some micro company users, so they have to consult an accountant for creating e-invoices. Again, the awareness in Latvia seems to be lesser than Estonia, it is moving forward faster.

Human mistakes are known to happen in the e-invoicing realm because there is a manual intervention. One example is that the e-invoice is sent to the wrong party. It is not a frequent problem but there have been instances.

E-invoicing roaming in the Baltics is built on peer to peer connections with agreements between the operators. Also, this is burdened by different web services for each operator. Regarding the content of e-invoices operators have their own validation rules and standards, even within a standard different version are used. A member of ITL (Estonian Association of Information Technology and Telecommunications) e-invoice working group pointed out that this problem is being analysed. It also emerged in several interviews that the status of the roamed invoice is not visible to all the parties while roaming.

E-invoicing receiving has issues such as having to always reconfigure the systems to receive e-invoices from different suppliers especially when the supplier count is high, some suppliers sending e-invoices through a centralized portal (in Lithuania), not having messages delivered upon receiving or not-receiving e-invoices and accepting wrong e-invoices or-invoices with errors.

Businesses are different from each other and it appears to be very hard to maintain one generalized e-invoicing system for all the businesses according to an expert, *"We still don't have a proper plug and play solution for e-invoicing because it requires a lot of development and implementation costs. The main reasoning is that the businesses are quite complicated and there are variations in the requirements such as Medicine, car Industrial, electronic industry, Banking and real-estate all these industries have their requirements. So, it's really hard to find what are the common requirements for everybody."* And it was also mentioned that each e-invoice operator has different systems.

Another major problem is the quality of data in e-invoices. This is a closely related issue with e-invoicing standards. Currently, there is a lot of non-validated data in the e-invoices and these can be erroneous. And different stakeholders interpret these data in different ways. There are non-mandatory fields in e-invoices which carry invalid data. An interviewee reported that "And sometimes the e-invoice receivers or the accountants have even declared not to use the e-invoices because we get a lot of invalidated data".

There are cost-related problems, especially affecting the small and micro businesses where the cost of connecting to an e-invoice operator and cost per e-invoice is high. This can affect the ERP partners as well, which was identified as a reason for the volume of B2B e-invoicing not developing as expected.

Business logic problems were also significant in e-invoicing as many participants of the study mentioned. Some of these are related to wrong tax calculations, lack of information on the e-invoice, erroneous values. One participant directly stated, *"even our company has some legacy software and, in some cases, the logic is not correct anymore."*

When talking about Latvia and Lithuania specifically, in Latvia and Lithuania the e-invoicing market model seems unsatisfactory. Centralized portals in Lithuania hinders fully automated e-invoicing processes and also B2B market restrictions. It works well with the B2G market. When compared to Estonia, Latvia and Lithuania do not have a central registry for companies to register for e-invoicing use. And the roaming infrastructure seems to be deficient. Another significant inadequacy in Latvia and Lithuania is the awareness and knowledge distribution compared to Estonia.

## 4.1.2 What is the current level of automation and what are the requirements for future automation opportunities in e-invoicing systems? (RQ2)

Table 7 Thematic Analysis Nodes and References for RQ2

| Sub-Theme | Number of Nodes | Frequency of references |
|---|---|---|
| Level of Automation | 1 | 6 |
| Automation Opportunity | 7 | 15 |
| General Requirements | 7 | 9 |

Throughout six interviews, the number of nodes for each sub-theme and the frequency of references to each category (node) are depicted by the above table.

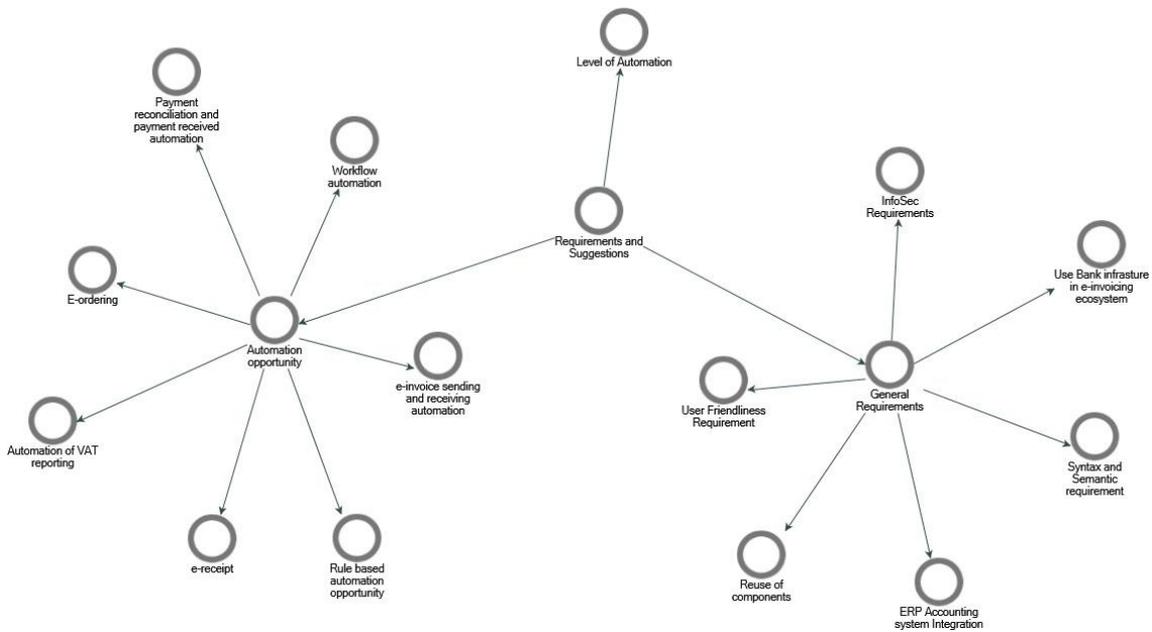

Figure 7 Automation Now and the Future (Thematic Map)

The level of automation in the Baltics is in a progressive mid-level according to experts. E.g.: *"It's not yet fully automated. It's kind of mid-level that you are in the direction of automation. There are some ideas. There are some solutions. We also tried to invest there and then put some effort there to develop or find some solutions."* E-invoice sending and receiving is quite automatic, but the processes around it need more automation. *"If you take the Estonian market again, it is a very good example that three operators, Omniva, Telema, Fitek, have all those kinds of automated value-added services such as Telema's*

*E-flow solution. They all provide some level of automation."* With this example, we can see that some solutions are already there in the Baltic region.

There were quite a few automation opportunities and general requirements identified from the qualitative data set.

Automation opportunities are listed below.

- Automated VAT reporting
- E-ordering
- E-receipt
- Payment Reconciliation Automation
- Payment Automation
- E-invoice workflow automation

Many of the automation requirements were identified by the researcher as 'Rule-Based Automation' which were, pre-posting of the documents, matching the invoices with orders, assigning workflows to an e-invoice, and automated journal entries for recursive invoices.

E-ordering was pointed out as an enhancement which proactively contributes to the e-invoicing and real-time economy realm. *"Because today we are kind of reacting on already closed cases. Invoice is mirroring something that has happened in the past. And if you want to be more efficient with handling the invoices, more automation, then it is better to have purchase order system in place while you already fix all your necessary parameters for this purchase order, which means that these carry on to the invoice and the invoice can be handled 100 per cent automatically because you have already approved the purchase ordering in the first place. I think this might be the direction that companies will be looking at in the future."*

As e-ordering has very special benefits, e-receipt was brought out to be the following step after e-invoicing, because it uses the same semantic and technical standards which is also an example from Finland. Experts have stated that payments automation and reconciliations of payments received are also the next steps in automation.

E-invoice workflow automation is already available with some e-invoice operators up to some level but they pointed out that some functionalities are yet to be implemented *"For example, as an e-invoicing service provider we don't have parallel workflows in approving e-invoices in our system. There can be very long invoices, maybe Telecom or real estate rental and 100 people should accept this invoice, and if you try to do this in serial it will last for a very long time."*

The general requirements consisted of,

- ERP and accounting system integrations

- Information security requirements which include software quality standards

- Reuse of components such as available payment infrastructure in banks

- E-invoice syntax and semantics validations

- Integrating of bank system infrastructure in e-invoicing

- Usability Requirement especially benefiting micro-companies

- Requirements which paves the way for the real-time economy in future

Estonian Ministry of Economic Affairs has a real-time economy working group specializing in integrating the vision of real-time economy according to an expert *"I know that in Estonia especially we have invested in the real-time economy in our Ministry of Economic Affairs, has this special working group"*.

## 4.1.3 What emerging technologies can be used to enhance the e-invoicing solutions eliminating the inefficiencies? (RQ3)

Table 8 Thematic Analysis Nodes and References for RQ3

| Sub-Theme | Number of Nodes | Frequency of references |
|---|---|---|
| AI | 2 | 18 |
| Blockchain | 1 | 8 |
| RPA | 1 | 9 |

Throughout six interviews, the number of nodes for each sub-theme and the frequency of references to each category (node) are depicted by the above table.

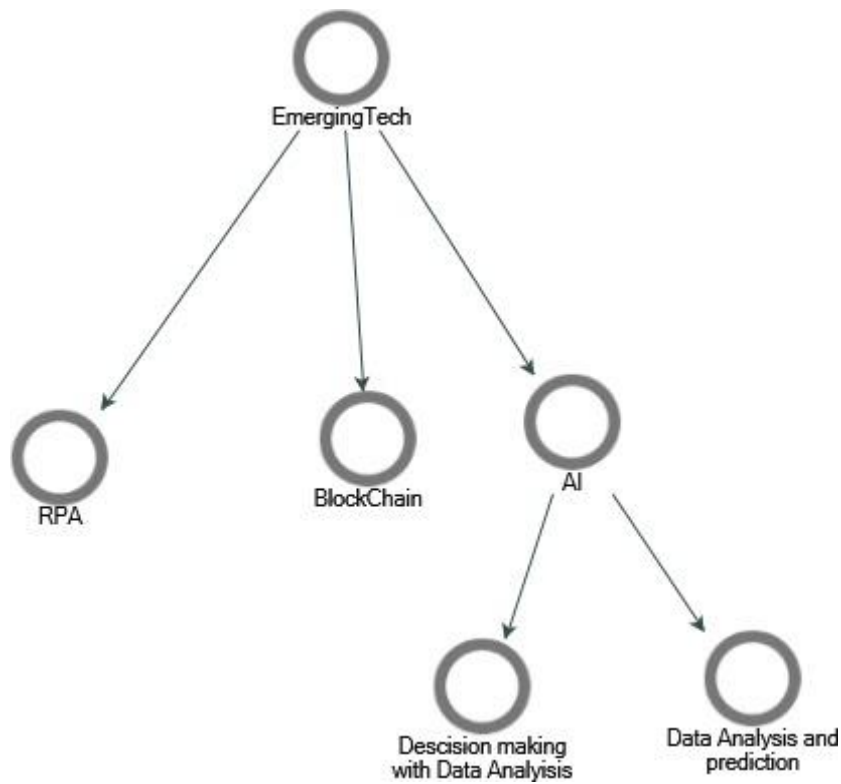

Figure 8 Emerging Technologies (Thematic Map)

Artificial Intelligence

Artificial Intelligence (AI) is considered to be implemented by several e-invoicing operators in the Baltics but not yet executed.

It emerged that AI can be used to

- Analyse the e-invoices and get the necessary information as a solution for e-invoicing standard issues (e-invoice receiving)

- Checking data quality

- Analyse and compare accounts posting templates

- Automating accounting entries

- Data Analysis and predictions

Most answers were towards using AI in purchase invoices, not sales invoices. And the challenges such as training the AI and making proper data sets for training, cost of implementing AI, trust in AI was discussed by experts.

Blockchain

As per the experts on e-invoicing, the blockchain technology can be used to

- Storing e-invoices, therefore, brings data integrity and document preservation

- Storing the timestamps of e-invoice changes and statuses

- Blockchain technology has been considered by only one of the e-invoicing operators, but they have not gone forward with it because of challenges such as

- Cost of the technology

- Cost of in human resources to make the platform

- Cost in technical infrastructure as it requires a lot of computing power

- The governments in the Baltic market are not demanding the authentication and authorization requirements in an ecosystem

Robotic Process Automation (RPA)

RPA was not implemented by any of the companies but it is considered and the uses and challenges were discussed.

It emerged that RPA can be put to use to

- Insert data into e-invoicing systems
- Data matching
- Order matching with e-invoices
- E-invoice approval process automation
- E-invoicing formats translation
- Extracting data from several different systems such as banking system, webshop interfaces etc without integrating all these systems
- Preparing Orders

It was obvious that some sort of automation is present in the e-invoicing systems, mainly some automation scripts. The benefits of using RPA is identified by the e-invoicing operators and it was mentioned in the interviews. E.g.: *"And sometimes it is impossible to even talk to those information system providers who take care of some platforms which are used by our customers, for instance, if you use AliExpress for delivering goods and Amazon Marketplace for your webshop and payment services from all kinds of different payment service providers, then you want to integrate all this data and put it together with your purchase invoices or purchases and such. So, it is very important to use RPA in this kind of situation. But if you look only to make accounting entries of bookings, use correct debits and credits, then it might be easier to use specific filters and scripts for that. It's not about RPA as such."*

Attitude Towards Engaging Emerging Technologies in e-Invoicing

The ideas from the experts about using the emerging technologies were both positive and negative. According to the below table of sentiments (made with Nvivo12) expressed about applying emerging technologies, positive sentiments are more than the negative sentiments. The number depicts the number of references which had positive or negative sentiment. Different shades of blue depict the intensity of the frequency of sentiment occurrence.

Table 9 Emerging Technologies Sentiments Analysis

| Technology | Positive | Negative |
|---|---|---|
| AI | 11 | 4 |
| Blockchain | 4 | 4 |
| RPA | 7 | 4 |

As per the above table, it is evident that Artificial intelligence has the highest tendency of being implemented in the future. RPA also has many positive sentiments than negative. And the blockchain has both positive and negative sentiments alike.

As per the thoughts of the author of the thesis, it can be the reason that many stakeholders are not very familiar with the blockchain and RPA technologies, for them these technologies seem very far away from being implemented in the near future. On the other hand, some of the experts mentioned that the cost of these technologies is much more. But when compared with the benefits in the future, this cost may be recovered. Nevertheless, the feasibility of implementing the emerging technologies are suggested to be performed in the future under *Chapter 6.3* Since AI has been already implemented in real-life software systems and it is very popular there may be more positive sentiments.

## 4.2 Analysis of the Survey for e-invoicing System Users

### 4.2.1 Demographic Analysis

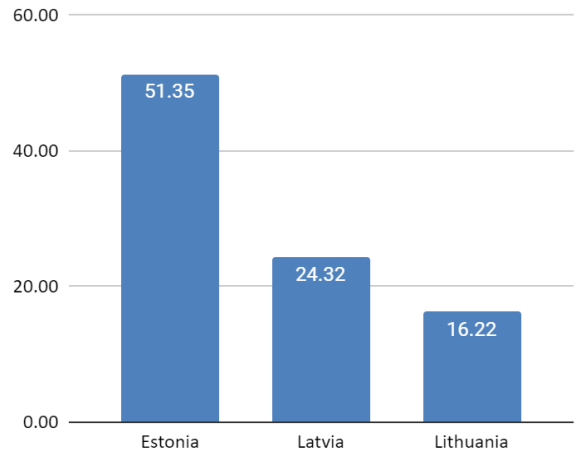

Figure 9 Respondent Representation of Companies Country-wise

To retrieve data for analysis purposes, a Google survey is built and shared among employees of organizations operating within the Baltic region. From the shared, 224 employees (varying from different designations) have responded based on their thoughts and feedback. As per the graph, it is clear that the maximum number of the respondents are from Estonia which is 114 respondents. Latvia had 54 and Lithuania had 36 respondents respectively. The reason for many users being from Estonia is that the market penetration of e-invoicing in Estonia is more than Latvia and Lithuania [25].

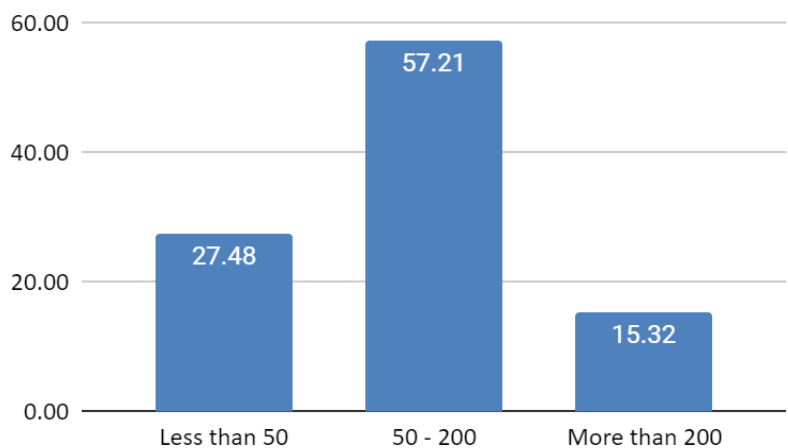

Figure 10 Organizational Magnitude in terms of Employees

Among the respondents, most of them are working in organizations having employees ranging from 50 – 200 which is more than 50% of the total respondents. Respondents working from organizations having less than 50 or greater than 200 are 27% and 15% respectively.

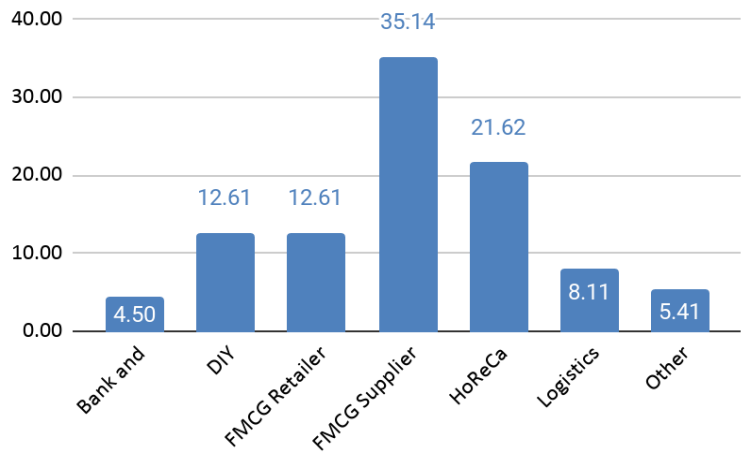

Figure 11 Industrial Sector-wise Respondent Representation

From the above graph, it is observable that most of the respondents working in organizations using e-Invoicing systems are FMCG Suppliers which is 35%. A considerable number of respondents (21%) are from the HoReCa sector which means hotels, restaurants and cafes. About 12% of the respondents are from FMCG retailer companies and another 12% of the respondents are from DIY companies.

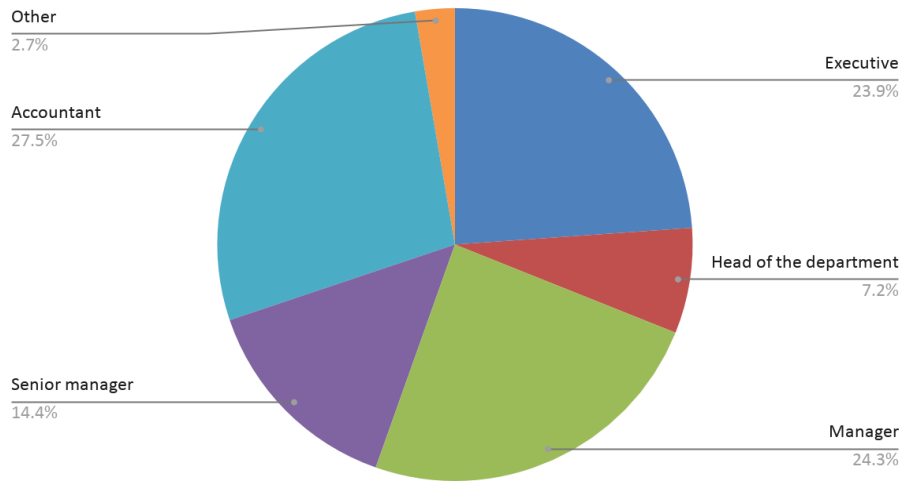

Figure 12 Respondents Representation by the Designation

According to the designation of the respondents, many are the direct and main users of e-invoicing systems who are accountants and executives. Also, huge portions are managers and senior managers because they approve invoices in the systems. Few heads of the departments have responded to the survey and others consisted of owner, CTO, software engineer designations. The author of this study thinks the user perception has originated from many different levels of designations at a company fairly divided, therefore all the functions accessible to all the designations are covered.

**4.2.2 User Perceptions towards e-Invoicing**

Perceived ease of use/ Effort Expectation

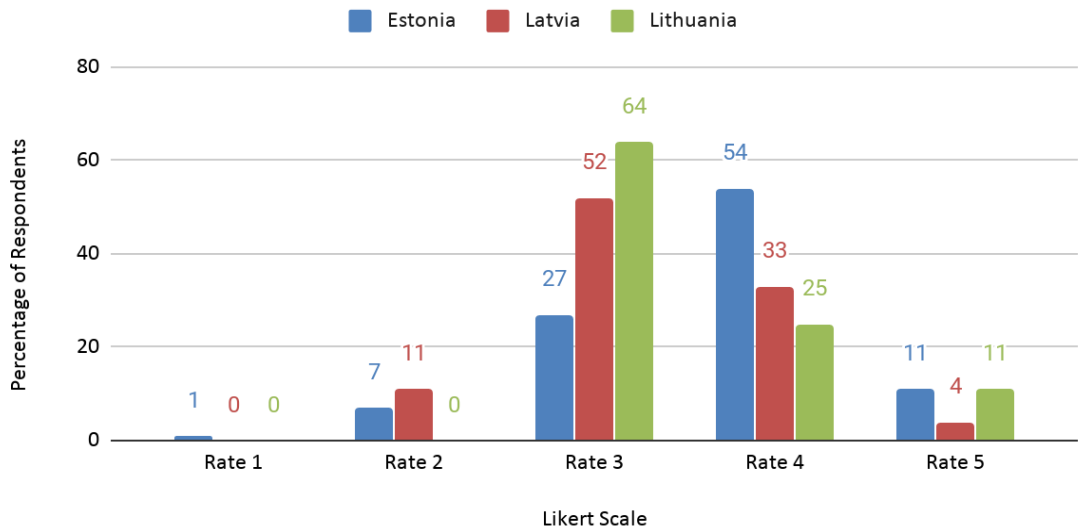

Figure 13 Perceived Ease of Use/ Effort Expectation in Estonia, Latvia and Lithuania

Regarding their thoughts given on Likert Scale ranging from 1 – 5 where 1 is least effective and 5 is very effective and user friendly, their feedback representation is as follows.

In Estonia, more than 60% among the respondents have rated 4 and 5 which illustrates that they are considerably happy with how effective and user-friendly their current e-Invoicing system is. 27% of respondents have rated 3 which reveal that they neither agree nor disagree with how effectively the current e-Invoicing system performs. Less than 10% only responded to unhappy experiences. Thus, it can be concluded that most of the respondents agree with current system performance and user-friendliness through certain improvements can yet be implemented.

In Latvia, the user perceptions are lesser than Estonia, and in Lithuania, it is even less. In Latvia and Lithuania, 52% and 64% of the respondents respectively have rated 3 reveal that they neither agree nor disagree with how effectively the current e-Invoicing system performs and 33% and 25% of the respondents respectively have rated 4 and they are somewhat happy with the e-invoicing systems. One of the reasons may be less awareness and experience in using more e-services with compared to Estonian users.

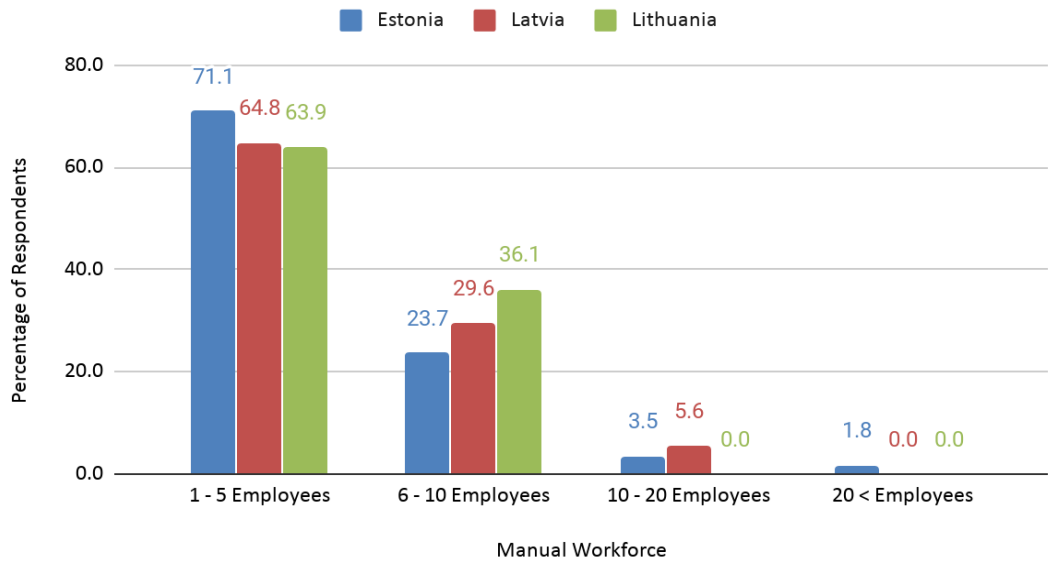

Figure 14 Manual Workforce on e-Invoicing Operations

From this question, the researcher expects to observe how much manual labour force is used among respondents' organizations for invoicing operations whilst having an invoicing system.

It is observable that most of the respondents' organizations use minimal labour force to get involved in day-to-day invoicing operational activities in all the countries but it is not zero. Around 64%-71% of the respondents' organizations use a labour force up to 5 people for e-invoicing operations while less than 5% of the respondents represent organizations using over 10 employees for manual invoicing activities. Hence it is safe to state that the current automation level on invoicing systems is at a moderate level and there can be more improvement.

Perceived Usefulness/ Performance expectation

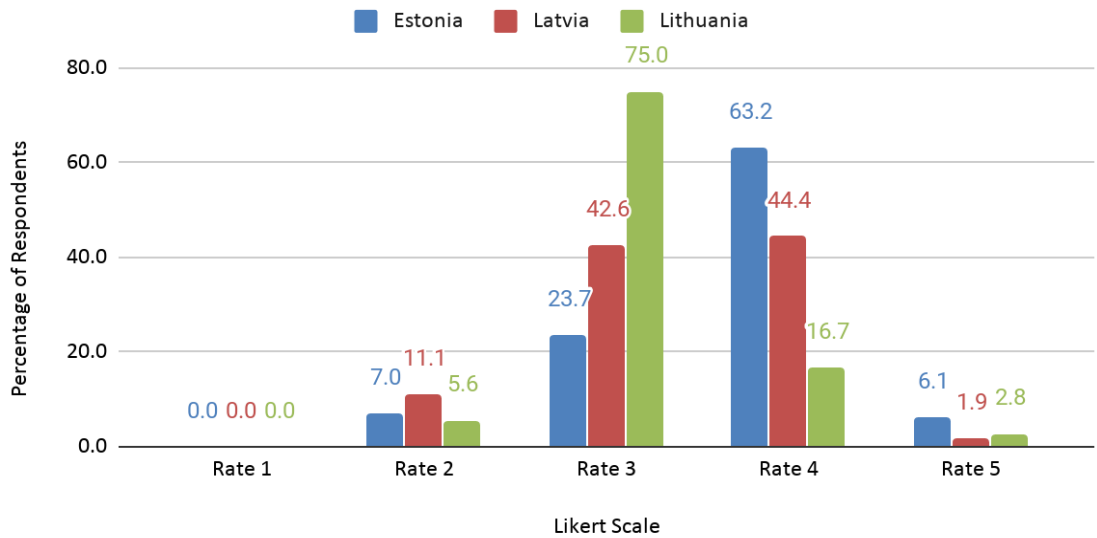

Figure 15 Coverage of Concerns Related to day-to-day e-Invoicing Operations in Estonia, Latvia and Lithuania

From the feedback received, analysis is conducted via a Likert scale from 1 – 5 where 1 represents lesser coverage while 5 representing higher coverage.

In Estonia, Most of the respondents (above 65%) agree that their current e-Invoicing system covers most of their day-to-day invoicing operational activities while 24% of respondents neither agree, nor disagree with the current functionalities. Only 7% express their disagreements on how useful the current features exist in their invoicing systems. Upon this observation, it can be concluded that most of the invoicing systems being used in Estonia do cover most of the features/functionalities, yet there is room for improvements.

In Latvia, it is observable that around 45% of the respondents vote in favour of current system functional coverage whilst another 40% - 45% of respondents neither agree to that statement nor do they disagree. With the 11% of respondents voting otherwise, it is safe to conclude that though a certain population among the respondents disagree with their system functional coverage, yet a considerable set of respondents do embrace the current feature set positively leaving room for further improvements in the current invoicing systems.

In Lithuania, most of the respondents (above 70%) neither agree nor disagree that their current e-Invoicing system covers most of their day-to-day invoicing operational activities. Among respondents, only 20% accept the fact that their invoicing system does provide their expected functional coverage. Only 5% express their disagreements on how useful the current features exist in their invoicing systems. Same as system effectiveness, compared to Estonian and Latvian respondents, Lithuanian respondents falls more towards disagreement partition, yet it is safe to declare that over 50% of respondents still are in favour of positivity of the required coverage in their invoicing systems.

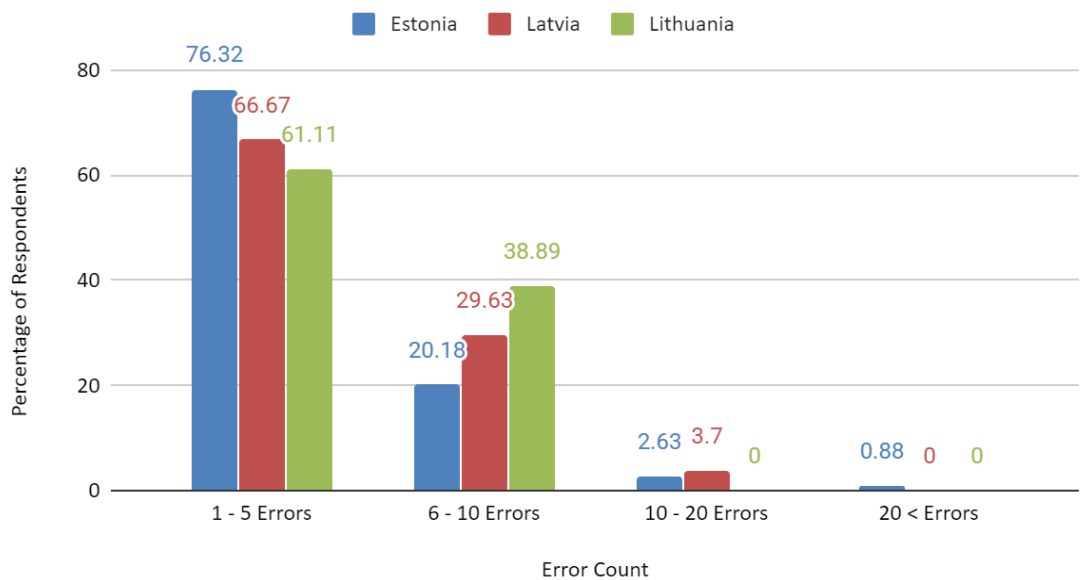

Figure 16 Daily Error Count in e-invoicing Systems in Estonia, Latvia and Lithuania

This is another input from respondents to analyse how effective the current e-invoicing systems are. It is visible that most of the invoicing systems have about 1-5 daily errors on average. More than 90% of the respondents agree that their respective e-invoicing systems or operations produce an average of fewer than 10 errors. There is no measurement of whether these errors are minor or not. Nevertheless, it can be assumed that with the need for higher automation coverage and lesser labour force, daily error count can be progressively reduced.

Perceived Risk

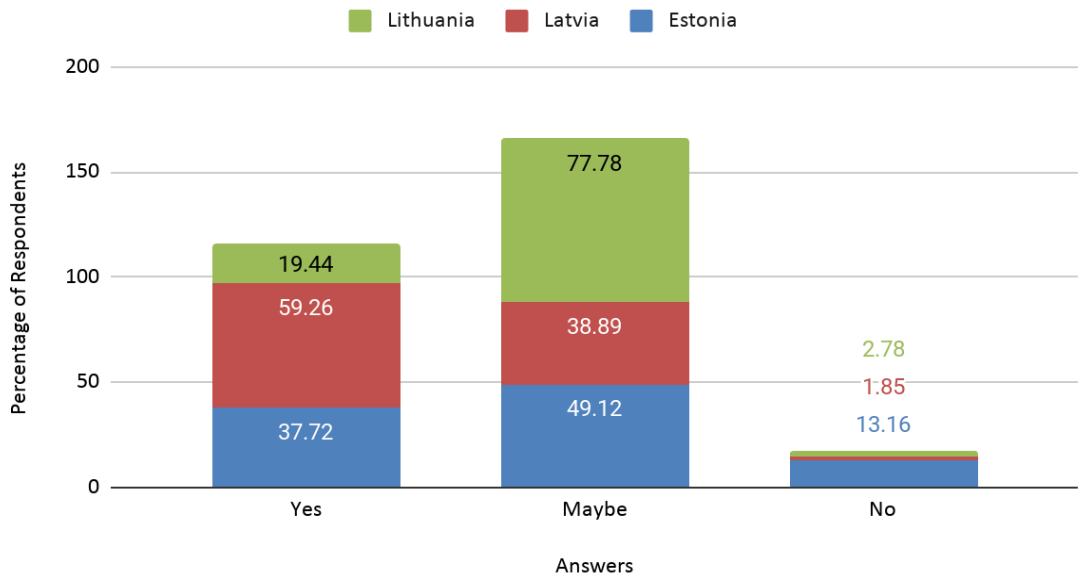

Figure 17 Perceived Risks in e-invoicing Systems in Estonia, Latvia and Lithuania

Under this question, the e-invoicing user's perception of risk in e-invoicing systems was asked in a manner that whether the users think there can be many risks. In general, there were many 'May Be' answers which shows that users have a doubt or they simply do not know. It can be stated as they are not denying the fact that there can be risks related to e-invoicing solutions. Also, a fair percentage of users said that there are risks. Only a very few said there are no risks associated with e-invoicing systems.

When we take the Baltic countries separately, nearly 50% of respondents said there 'maybe' risks, and 38% said there are risks and 13% said there are no risks. In Latvia, 60% of respondents said there are risks definitely, and 38% said there may be risks. In Lithuania, 77% of respondents said there may be risks and only about 20% said there are risks. 1%-3% Both Latvian and Lithuanian respondents said no for an answer.

This means most of the respondents know that there can be risks involved in e-invoicing systems and it can be a good thing that this is known and mitigations or improvements can take place.

Perceived Trust

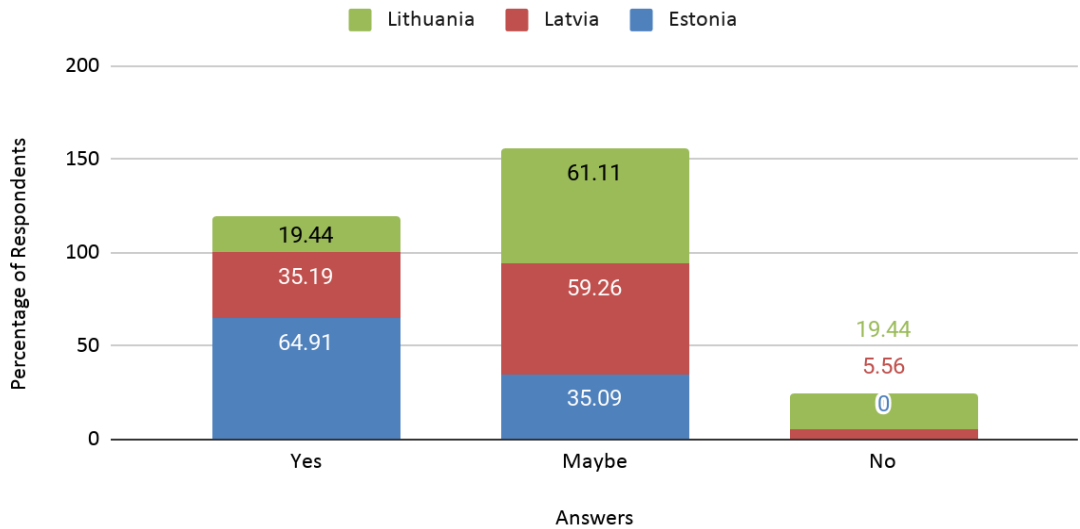

Figure 18 Perceived Trust in e-invoicing Systems in Estonia, Latvia and Lithuania

Under these answers to the question, the perception of trust related to e-invoicing solutions is measured. When taking the Baltic region in general, a lot of answers contain 'maybe' which means the users are not sure where they trust the e-invoicing solutions or not. But when we take the percentage of users who have said 'yes' to the above question, it means they trust e-invoicing solutions. Very few percentages of the users do not trust. We can conclude that most of the Baltic users are positive on the perception of Trust.

When we compare Estonian users with the Latvian and Lithuania users, more than 50% of the respondents (half of the surveyed population) trust e-invoicing systems. This can be the case that Estonia is a more mature e-state than the rest of Baltic countries. In contrast, nearly 60% of the Latvian and Lithuanian respondents are not sure whether they trust e-invoicing solutions. But definitely with the enhancements and automation the trust perspective can be brought towards a 'yes' answer in the future.

Perceived Information Security

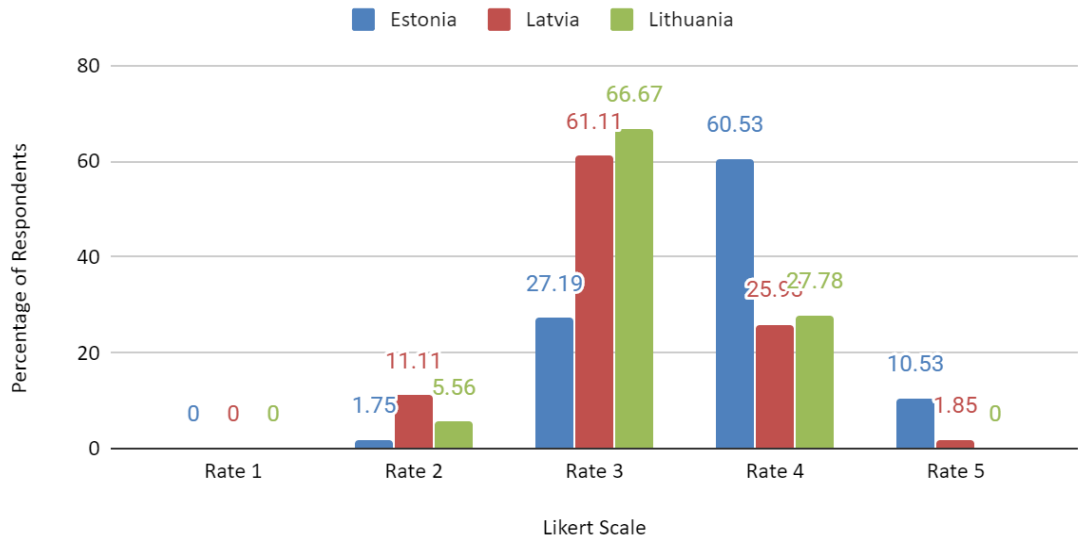

Figure 19 Perception on Information Security in e-invoicing Systems in Estonia, Latvia and Lithuania

From the feedback received, analysis is conducted via a Likert scale from 1 – 5 where 1 represents least information security and while 5 representing the most information security.

More than 60% of the respondents from Latvia and Lithuania think that the e-invoicing system is fairly secure, while about 25% think there is good security. Almost none think either it is very poor or excellent in information security. When we consider the respondents from Estonia, the majority of 60% thinks that information security is good in e-invoicing systems and 27% of the respondents think information security in e-invoicing systems is just fair. We can especially notice that 10% of the Estonian e-invoicing users think that the information security aspect is excellent in e-invoicing. This can be due to the mature e-invoicing systems and e-invoicing operators in Estonia. In fact, in Estonia, most of the government is digitalised and documents in public agencies are digitally exchanged [84]. Therefore, is it possible to increase the level of trust in the same way in the e-invoicing domain too.

Automation Requirement on e-invoicing

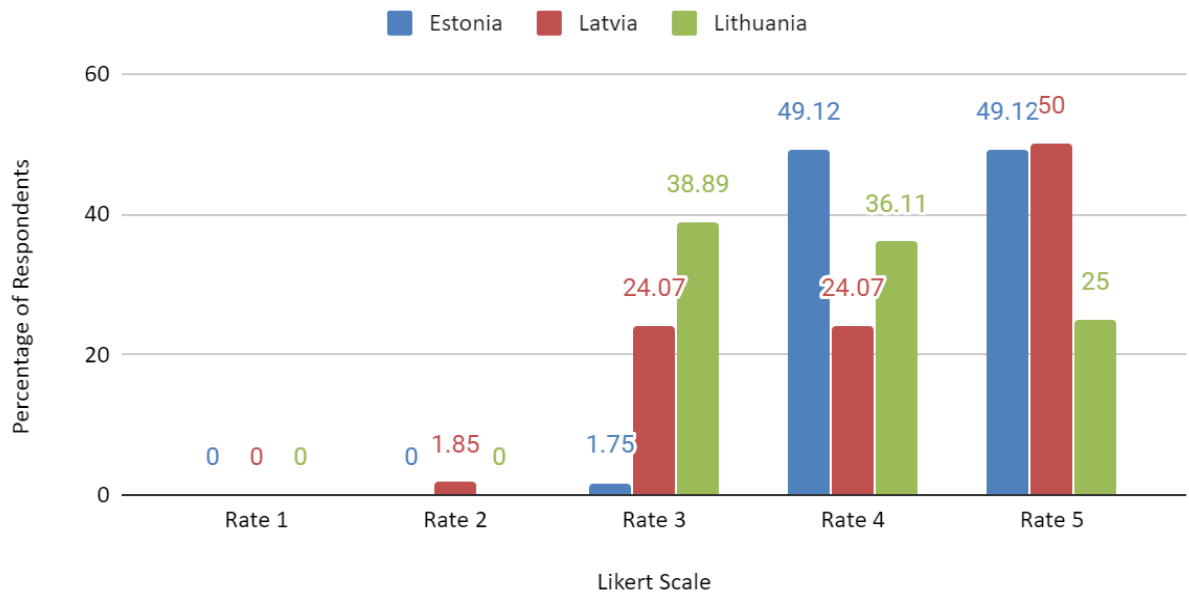

Figure 20 The Requirements of Automation in e-invoicing Systems in Estonia, Latvia and Lithuania

From the feedback received, analysis is conducted via Likert scale from 1 – 5 where 1 represents that the user does not need any automation at all in the e-invoicing process while 5 represents that everything should be fully automated which can be called as touchless e-invoicing.

When we consider Estonian e-invoice users who responded, 50% of them would like to have fully automated systems while 50% of them would like to have nearly automated systems where they have manual control over some aspects.

In Latvia again similar to Estonia, 50% of the users would like fully automated systems, where 25% of them would like to have half automated and 25% of them would like nearly automated e-invoicing systems. In Lithuania, only 25% of the respondents would want to have a fully automated system, and about 39% of the respondents would like to have half automated and 36% of the respondents like to have nearly automated systems.

There may be various reasons some users do not like fully automated systems. The author thinks one of them is that they might be thinking that they will lose control over the systems.

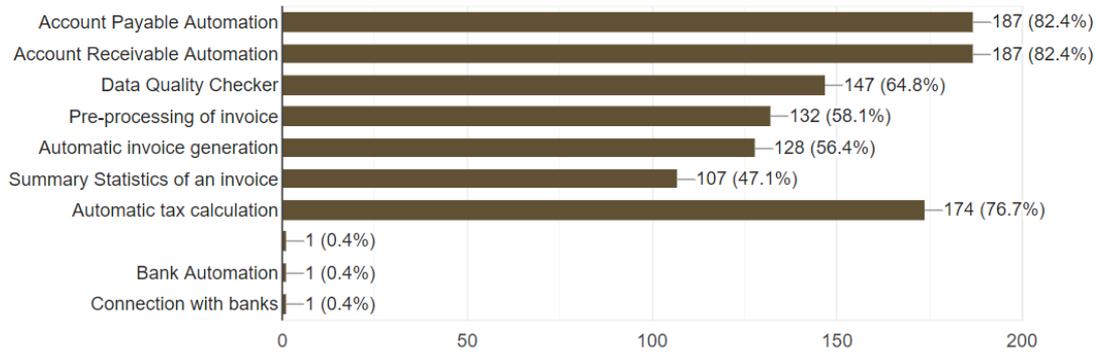

Figure 21 Features Requested by the Survey Respondents

These additional services were included in the survey and users' requirement of having the same functionalities built into their e-invoicing systems were asked to be marked. It is clear that over 80% of respondents request for account payable and receivable automation followed by automatic tax calculation at 76% and data quality checker at 65% respectively. 58% of the respondents would like to have pre-processing of the e-invoices automated and 56% of the respondents would like to have invoices created automatically. Also, 47% of the respondents would like to have summary statistics displayed for an e-invoice, especially when the invoice is too long containing many invoice rows. By implementing the same requirements on current invoicing systems, it can be stated with confidence that it will have a positive impact on current frustration respondents/users face during invoicing processes. Also, since this proves users' requirements, it shows the tendency of possible usage of the functionalities eagerly by the users in the Baltics.

# 5 Recommendations

This chapter brings out the recommendations by the author of the thesis as a suggested ecosystem with Emerging Technologies for e-invoicing. The benefits of the suggested eco-system are explained which answers the problems listed under the 1st research question and suggestions derived from 2nd and 3rd research questions. Finally, the limitations of the research are produced.

## 5.1 Suggested Software Ecosystem with Emerging Technologies

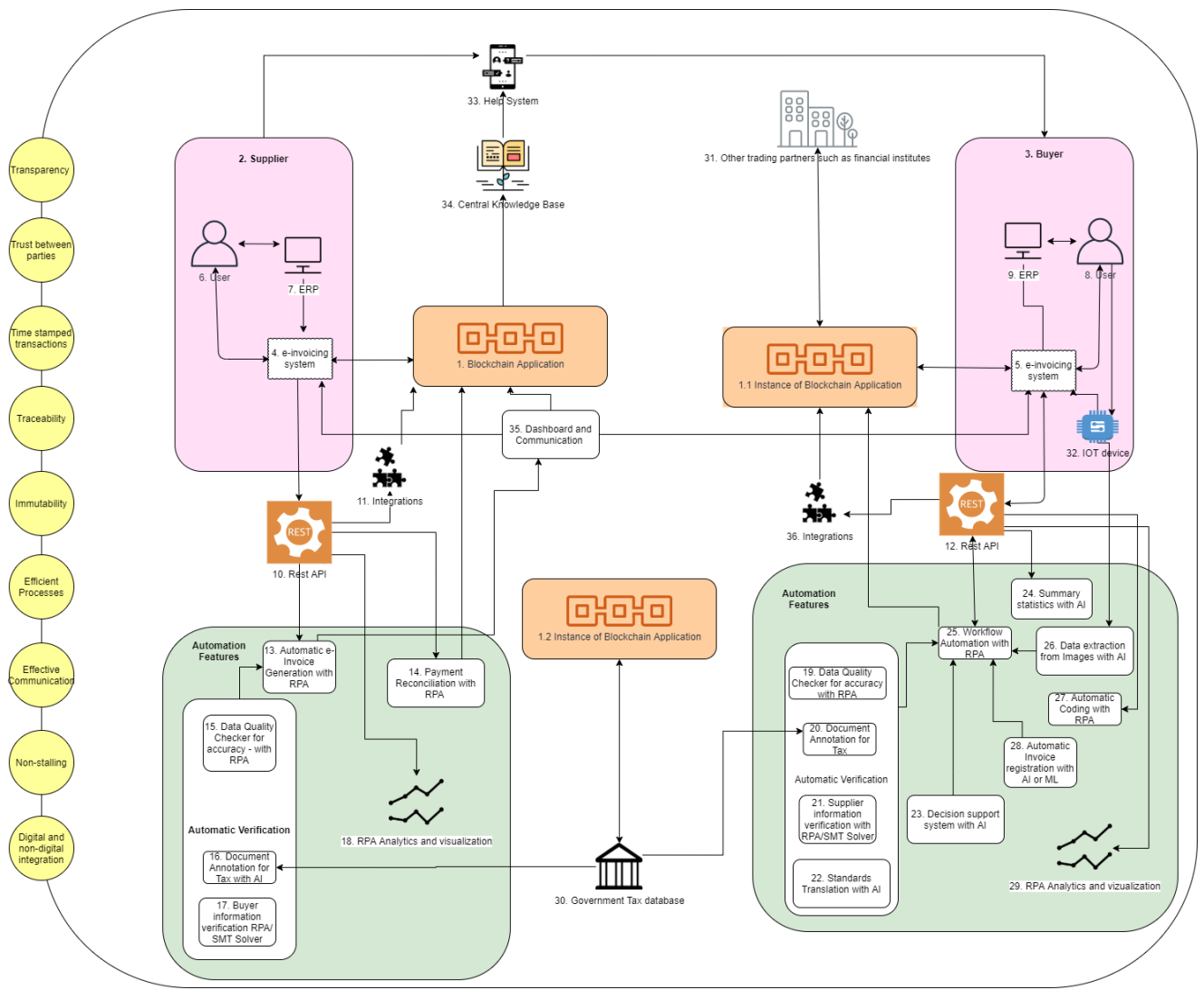

Figure 22 Suggested Software Ecosystem with Emerging Technologies

This design is a blockchain-based ecosystem where suppliers, buyers, trading partners such as financial institutions and government institutions such as tax authorities are the entities that participate in the network. In the diagram only one entity from each supplier, buyer etc is depicted, but in reality, there are many instances of the blockchain where many participants can join. The type of blockchain network is advised to be private or consortium since the entities who can join should have the consent to join and participate. Eg: A new supplier should have the necessary permissions and agreements with the service provider and buyers he is associating with. Therefore, this model is furthermore centralized than public blockchain.

Blockchain registers transactions when,

- E-invoice is generated by the buyer

- E-invoice is registered by the supplier

- E-invoice processed through the workflow (can have several transactions registered according to the requirement)

- Tax is registered with the tax office

- E-invoice is paid

- Other instances as the requirements arise

The items 4,6,7 shows a user from the supplier company is connected to their own ERP/accounting system and the e-invoicing system. The items 5,9,8 shows a user from the buyer company is connected to their own ERP/accounting system and the e-invoicing system. A user mainly uses the ERP system but there are user interfaces in the e-invoicing system for e-invoicing process related tasks. These two systems are separated because not all the ERP systems have e-invoicing functions available and it is easy to maintain. Just the integration between the two systems is necessary.

There can be different e-invoicing systems connected to the blockchain according to the e-invoicing model. The e-invoicing system/systems of the users are connected to the RPA, AI and other components with web services, which is depicted as the Rest API (item 10 and item 12).

With the item 11,36 - integrations are capable of the e-invoicing systems to be integrated into other software or tools as the requirements arise. Also, these integrations can have a transaction stored on the blockchain network, also under the requirements of each company/ organization.

When describing the item 13- automatic e-invoice generation which is an RPA service, it includes getting necessary data from the ERP of the supplier. 56% of the e-invoice users said they would prefer to have this functionality in an e-invoicing system. RPA is suggested in this case because invoice generation is rule-based, matched with an order. Also, RPA can perform collecting data from different systems and generating one document which answers the problems of using different systems by the company. While the e-invoice is getting generated, the verification on the e-invoice is done by the sub-module labelled as 'Automatic Verification'. This includes data quality checking with RPA (item 15), especially the syntaxes and semantics according to the required standard. Because different buyers can be using different standards as mentioned in the thematic analysis of this study. The rules for data quality checks should be identified before designing the functionality and could be enhanced with the usage. Another component of the automatic verification is document annotation for tax (item 16). The tax values can be calculated against the government tax database rules and this component should have a connection with the government tax database. The other component is buyer information verification (item 17), which can be done with a simple rule-based matching process with RPA or SMT solvers (Satisfiability Modulo Theories) which includes higher degree verifications [83].

The payment reconciliation of e-invoices (item 14) is done according to the matching with payments received against an e-invoice. This is a simple rule-based process.

With RPA analytics and visualizations (item 18), there can be a dashboard of summary statistics of invoices, taxations, goods, buyer related information etc. During the study, 47% of the e-invoicing users stated that they would prefer to have this feature of summary statistics.

Item 26 which is Data extraction from images with AI, this can include PDFs and other non-machine-readable formats. In an e-invoicing system, an image or a PDF is not counted as an e-invoice but there is some percentage of non-readable invoices still circulating in the Baltic region. AI is capable of training to read these invoices and convert them to an e-invoice until all the B2B invoices become fully electronic. This can have inputs from an IOT device where the users are out of office scenarios with the integration of item 32.

Summary statistics with AI is used to show the buyer a condensed summary of very long invoices. This can help to make decisions without having to read the whole invoice. As an example, the invoice for electricity in different locations of a huge retail chain.

Under the automatic verification of the buyer's side, data quality checker with RPA (item19), document annotation for tax (item 20), supplier information verification with RPA/SMT Solver (item 21) is the same as supplier's automatic verification module. The speciality here is there is a functionality of standards translation using AI (item 22). This is the solution for significant standards related problems emerged in the study where there is no proper standard and different versions of the standards are being used. With AI, there is no requirement to have one proper standard instead, each standard can be analysed and translated into other standards as required.

Item 28 which is Automatic invoice registration to the ERP or accounting system can be done with AI or machine learning. It can be trained with the exceptions for each supplier etc. Also, this feature can include the invoice returning and so on.

e-Invoicing workflow is automated with RPA (item 25) as it is a rule-based process and this is complemented with a decision support system with AI (item 23). The decision making can be either the decision to be taken by AI or the decision can be prompt to the user, so that the user can either go ahead or not in the workflow.

Automatic coding is simply another rule-based process which is automated with RPA where the accounting entries are made to the ERP or accounting system.

As same as item 18 in the supplier's side, item 29 is RPA analytics and visualization for the buyer concerning all the e-invoices which were again a required feature to make informed decisions.

Dashboard and communication (item 35) is a common functionality which is integrated into the whole ecosystem to answer the question of visibility. This is an inefficiency mentioned by many experts during the study that buyers, sellers and e-invoicing operators do not know the status of an e-invoice or the ability to send e-invoices across the Baltic region. With the data in the blockchain, this visibility is made possible.

The help system (item 33) which is connected to a central knowledge base (item 34) is the presence of a chatbot to enhance the communication between trade partners. The main reason this functionality is different to the dashboard is that dashboard is a dynamic information display related to e-invoices, where the help system is capable of answering queries on trade-related matters. This was also brought up by an expert during the interviews that the communication between trade partners regarding e-invoicing abilities is very hard. As an example, a supplier can ask the help system to retrieve the agreements that need to be re-evaluated at the end of a period. The help system ought to be designed in a way that each participant has their limitations of accessing information.

## 5.2 Expected Benefits of the Proposed System

As depicted in the ecosystem diagram, there are several benefits of the proposed system, which also answers some of the problems identified in the thematic analysis of this study. Individual benefits of each component of the proposed system are discussed under the topic 5.1.1.

Transparency - Using a blockchain-based ecosystem gives transparency to all the parties where it is auditable with valid ledger transactions. Current e-invoicing systems in the Baltics are heavily dependent on the e-invoicing providers and this can be solved with the use of blockchain. Also, it minimizes the cost of maintaining contracts and reduces the risks of operational risks [82].

Trust between parties who trade - With the blockchain architecture, it needs consensus to be on a node which is an agreement between all the participating parties or relevant parties [82]. Even smart-contracts can be involved; hence all the trading partners are under the agreements which enhances the trust. This also brings the customer perception of trust to a higher degree.

Time-stamped transactions - Blockchain offers time-stamped transactions [13], therefore it is kept up to the agreements between trade partners about the number of days for payments etc. This also helps audit controls and conflicts management.

Traceability - Since the transactions are recorded in the chronological order and it is replicated in many copies at the distributed ledger across all the nodes [13], the transactions are easily traceable.

Immutability - The information stored on the blockchain remains unchanged which means the information is indelible as well. This is also one of the noteworthy features of using blockchain especially in the financial sector [82]. This contributes to making the user perceptions very effective towards trust and security.

Efficient processes- Current e-invoicing systems in the Baltic regions are either private systems or directly connected to trade partners or connected through an e-invoice operator. Therefore, the processes are inefficient in many ways as discussed in the 4th chapter. With RPA processes for manual processes and involvement of AI can bring a lot of efficiency in e-invoicing. The survey revealed that manual effort is consumed in the companies for e-invoicing processing which also requires a substantial amount of time and is prone to errors. With well-trained AI and properly designed rule-based RPA processes can make e-invoicing efficient to a greater degree.

Effective Communication - The communication between trade partners is made effective with the Dashboard and communication feature which is also receiving data from blockchain transactions. This is a solution to the problem discussed in the analysis of this study, where trade partners do not know the status of an e-invoice. Also, the help system which is connected to the central knowledge base (which can be a government registry also) can provide information to all the trade partners acting as a chatbot where effective communication is benefitted. Also, with all the information existing in the blockchain network, the communications between trade partners can be effective because all the information is available and accessible by the users.

Non-Stalling - This is an added advantage of using automation features where the e-invoices do not get stalled at a trade partner since there is minimal manual intervention.

Digital and non-digital integration - Data extraction with AI and incorporating IoT devices in the ecosystem makes non-digital aspects to be digitized. Because still there is a considerable usage of non-machine-readable invoice usage in the Baltics according to the study.

This kind of software ecosystem will be making sure that a tight collaboration is present between the public and private sector. An as an example, how Estonian data exchange layer assisting private-public partnerships[85], the same way e-invoicing ecosystem also can help strengthen private-public partnerships with the government authorities participating in the blockchain network.

## 5.3 Limitations of the Research

- The sample selected for the study represents the users and experts of e-invoicing systems in the Baltics. But the users who answered may not be the voice of all the users in Baltics and the information collected through the survey and interviews may not be representing every e-voicing user and expert in the Baltics. In the study, we generalize the data for the whole population.

- Qualitative data gathered from the interviews are limited to 6 interviews. Qualitative data is said to be limitless in being gathered and analysed. Here in the study with the limited time frame, only the data which needs to answer the research questions are studied further.

- The views of ERP Partners and other integrated systems and stakeholders on e-invoicing could have been taken into account to gain more insights, but again with the master thesis scope and time frame, this was not possible.

- The recommendations eco-system is a suggestion. The eco-system may be impossible to be fully implemented in the current situation with the limitations such as cost and magnitude of the recommended system, but there can be some parts of it which are feasible to implement later paving the way to full implementation.

# 6 Summary and Suggestions for Further Research

This chapter brings out the summary of findings from the research study. Finally, further research suggestions are produced by the user.

## 6.1 Conclusion

The study concludes that in the Baltic region there are inefficiencies in e-invoicing systems and there is room for improvement. The perceptions of users regarding e-invoicing systems are moderate or fairly positive. There is a potential to use emerging technologies to minimize inefficiencies and introduce more automation to the e-invoicing systems. The proposed e-invoicing ecosystem can be researched further in the Baltic region as a next step.

## 6.2 Summary of Key Findings

RQ1:

During the study, it was identified that there are several inefficiencies in the current e-invoicing systems in the Baltic region. Most of them are operational inefficiencies where the e-invoicing standards-related issue was the most discussed. It was closely related to other operational issues such as business logic problems, not supporting different types of businesses and quality of data. Common problems in the Baltic region consisted of cost-related issues in implementing e-invoicing solutions, integrating ERP systems, human errors in e-invoicing, small or micro companies are not being taken into account by major e-invoicing operators, PDF and paper invoice usage, e-invoicing receiving problems and e-invoice roaming issues. There were specially identified problems, in Lithuania the inefficient e-invoicing market model and less e-invoicing awareness in Latvia and Lithuania both. Other two types of inefficiencies included information security-related problems and technological problems. The most significant issues were identification, authentication and authorization of e-invoices and trade partners. Relying on e-invoicing operators for information security was noteworthy in the Baltic region, especially in Estonia. The technology is old and problems in legacy software were also brought into attention by many interviewees.

Survey demographics showed that most of the users who identified are from Estonia, who work in the middle-level organizations of the magnitude, mostly from FMCG and HoReCa sectors. Users have many different designations within the company so the author believes that a fair amount of answers carry all aspects of an e-invoice system.

Perceived ease of use or effort expectation on e-invoicing systems is mostly neutral while about 40% of users have a perception towards slightly effective. It was evident that most companies have up to 5 employees and some companies have up to 10 employees doing manual functions in the e-invoicing systems which implies there is room for automation improvement. Perceived usefulness of e-invoicing systems or performance expectation is mostly neutral while about 40% of users have a perception towards slightly effective. Majority of the users are facing up to 5 errors and some users are facing up to 10 errors daily. This could be one of the reasons that the users' perception is not very high. The perceived risk factor is also neutral in general while in Latvia perceived risk is higher compared to Estonian and Lithuania. Perceived Trust is also Neutral in general but in Estonia, perceived trust is higher comparatively. The reason for this can be the maturity of e-services. The perception of information security is neutral as a region, but it is more towards fairly effective in Estonia. Again, the reason could be that Estonia is a mature e-State.

RQ2:

The level of automation present in the e-invoicing systems in the Baltic region was understood to be to a certain degree. Some efforts have been incorporated towards automation and there is a vision for a real-time economic future in Estonia. The study brought out automation opportunities not only in the e-invoicing process but other highly intertwined processes with e-invoicing such as e-receipts, e-ordering, payments reconciliation and VAT reporting. Rule-based automation prospects were also discussed which can be easily automated in the future. From the general requirements, using the already existing bank and payment infrastructures were brought out and also the advantages such as the re-use of components. General requirements consisted of usability, information security, ERP partners integration etc. These requirements and opportunities were discussed at the surface level, but there can be many more underlying requirements when conducting in-depth requirements gathering for an implementation.

Concerning the survey, the request for automation in e-invoicing systems is generally very high in the region, where most of the users' requirement is fully automated touchless e-invoicing or fairly automated e-invoicing. The automated features requests include account payable automation, account receivable automation significantly. Other features such as data quality checker, automatic tax calculation, automatic invoice generation and pre-processing of the invoice are also in demand.

RQ3:

Using emerging technologies for the advancement of e-invoicing has mixed opinions from experts. But the positives outplayed the negatives when the sentiments were analysed in the qualitative study. The biggest concern was the cost and effort to implement emerging technologies such as AI, blockchain and RPA. Many experts pointed out that AI has a lot of potential in the e-invoicing future.

## 6.3 Suggestions for Further Research

The next stage is to conduct a feasibility study of the proposed eco-system for the Baltic region. Then it can be prototyped and tested paving the way for a real implementation if the tests bring positive results. It can be the whole ecosystem or a part of it and it can be implemented in stages.

The research findings relate mainly to the Baltic region. Therefore, the information this study produced is useful in making informed decisions by the stakeholders associated with e-invoicing.

The study can be reproduced in the European Union (in different member states), other countries or regions to discover inefficiencies in e-invoicing systems, automation opportunities with emerging technologies. Also, the perceptions of e-invoicing users can be useful when designing other e-services in the Baltic region, when applicable.

# Appendix 1 – List of Interviews

| Interview | Name of Interviewee (Experts/ Stakeholders) | Position |
|---|---|---|
| 1 | Bo Harald | Senior Advisor. Real Time Economy- RTE Program (Finland) |
| 2 | Andres Lilleste | Business Manager at Fitek who is a e-invoicing operator for 9 countries including the Baltics<br><br>Member of ITL e-invoicing working group |
| 3 | Toomas Veerso | Chief Technical Officer at e-invoicing operator Telema who is Pan Baltic e-invoicing operator<br><br>Member of ITL e-invoicing working group |
| 4 | Indrek Allas | Inbound Business Unit Manager at Fitek who is a e-invoicing operator for 9 countries including the Baltics |
| 5 | Margus Tammeraja | Chairman of the Board at Association of Estonian Accountants |
| 6 | Aleksandr Beloussov | Head of Business Reporting Systems Unit at Centre of Registers and Information Systems Estonia |

# Appendix 2 – Thematic Map of Qualitative Analysis

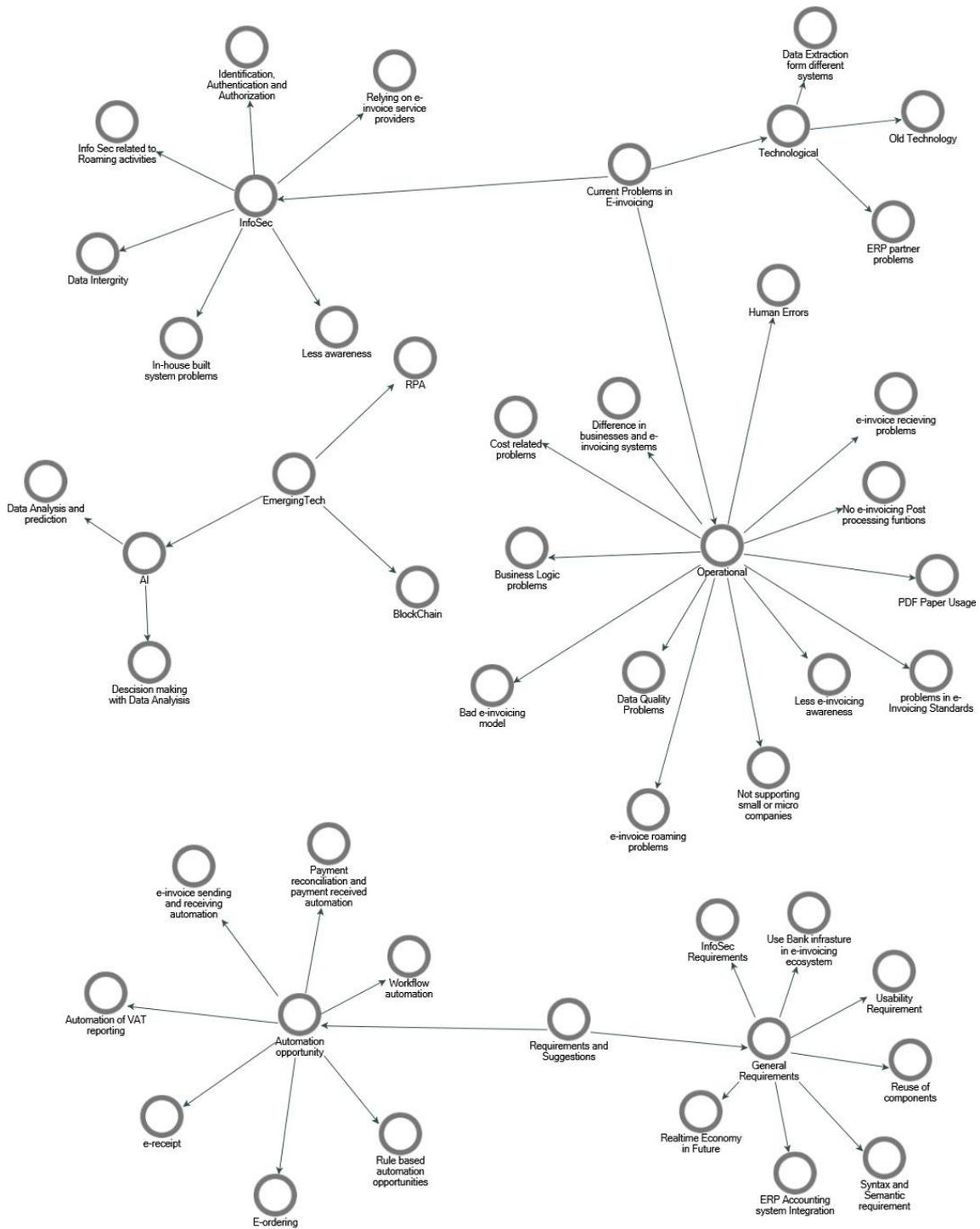

# Appendix 3 – Interview Questionnaire Outline

| Research Question | Key Element | Interview Question |
|---|---|---|
| RQ1: What are the current inefficiencies of existing e-invoicing systems? | Scope / Operational gap | How different the current system functionalities are from your expectations? Do you see major gaps among the system functional specifications? |
| | Performance | What's your opinion on the current level of performance of the system? (In terms of operational, user-friendliness, speed, etc...) |
| | Security | What are the major drawbacks you witness in the system with regards to information security with digital content within the system? |
| | Technological | What are the technological problems in current e-invoicing systems? |
| RQ2: What is the current level of automation and what are the requirements for future automation opportunities in e-invoicing systems? | Technological effectiveness | What kind of a technological stack is used for current automation processes in the e-invoicing system? How effective are the chosen tech stacks in automation? |
| | Functional effectiveness | How efficient is the current functional automation available in the system? |
| | Coverage of automation (room for improvement) | What do you think of the current automation coverage in the system? Do you think there's more room for automation improvement? |
| RQ3: What emerging technologies can be used to enhance the e-invoicing solutions eliminating the inefficiencies? | Technological gap (past vs current) | With the increase in technological growth / improvement, how efficient do you think the used technologies in the current e-Invoicing systems are compared to now-a-days improved technical ground? |
| | Inefficiencies in traditional technologies | What are the loopholes you can observe in traditional technologies that are used in current e-Invoicing systems? |
| | | What do you think about using blockchain in the e-invoicing process? |

| Research Question | Key Element | Interview Question |
|---|---|---|
| | Blockchain, RPA and AI effectiveness | What do you think about using Robotic Process Automation (RPA) in e-invoicing processes? |
| | | What do you think about using Artificial Intelligence (AI) in the e-invoicing process? |

# Appendix 4 – Survey Questionnaire Outline

| Research Question | key element | Survey Question | Question Type |
|---|---|---|---|
| Demographics | Region | Which country does your company operate in? | Selection of the country/countries (Option to write other is included) |
| | Company Size | What is the size of your company? | Three ranges of company size determined by the employee count <50, 50-250, >250 |
| | Position of the user | What is your current position in the company? | Selection of designation (Option to write other is included) |
| | Company Sector | Which sector/segment does your company belong to? | Selection of sector (Option to write other is included) |
| | Usage of Invoices | Which invoices do you use at your work? | Selection of options (Sales invoices, Purchase invoices, Both, none) |
| | | How do your company receive and send invoices? | Likert Scale of 0-100% for Paper invoices, PDF/Word invoices, e-invoices |
| What are the current inefficiencies of existing e-invoicing systems? | Perceived ease of use/ Effort Expectation | How effective and user-friendly the current e-Invoicing system is for you to conduct your invoicing activities? | Likert Scale of 1-5 (lease effective - Very effective) |
| | | How many people are working on manual tasks such as allocating costs/ approving invoices/ correcting data on invoices? | Short answer as a number (answer was validated to be a numerical value only) |
| | Perceived usefulness/ | Do you think the current e-Invoicing system covers all your concerns related to your | Likert Scale of 1-5 (Not cover at all- Covers everything) |

| Research Question | key element | Survey Question | Question Type |
|---|---|---|---|
| Demographics | Region | Which country does your company operate in? | Selection of the country/countries (Option to write other is included) |
| | Company Size | What is the size of your company? | Three ranges of company size determined by the employee count <50, 50-250, >250 |
| | Position of the user | What is your current position in the company? | Selection of designation (Option to write other is included) |
| | Company Sector | Which sector/segment does your company belong to? | Selection of sector (Option to write other is included) |
| | Usage of Invoices | Which invoices do you use at your work? | Selection of options (Sales invoices, Purchase invoices, Both, none) |
| | | How do your company receive and send invoices? | Likert Scale of 0-100% for Paper invoices, PDF/Word invoices, e-invoices |
| | Performance expectation | day-to-day invoicing operation? | |
| | | How many errors (data/document/system) do you encounter in your e-invoicing system per a day? | Short answer as a number (answer was validated to be a numerical value only) |
| | Perceived Risk | Do you think the current e-Invoicing is probable to be exposed to certain risk factors? (E.g.: technical, functional, operation, etc...) | Answer selection: Yes/No/Maybe |
| | Perceived Trust | Do you trust digital content? Do you think the current e-invoicing solution you are using is trustworthy? | Answer selection: Yes/No/Maybe |

| Research Question | key element | Survey Question | Question Type |
|---|---|---|---|
| Demographics | Region | Which country does your company operate in? | Selection of the country/countries (Option to write other is included) |
| | Company Size | What is the size of your company? | Three ranges of company size determined by the employee count <50, 50-250, >250 |
| | Position of the user | What is your current position in the company? | Selection of designation (Option to write other is included) |
| | Company Sector | Which sector/segment does your company belong to? | Selection of sector (Option to write other is included) |
| | Usage of Invoices | Which invoices do you use at your work? | Selection of options (Sales invoices, Purchase invoices, Both, none) |
| | | How do your company receive and send invoices? | Likert Scale of 0-100% for Paper invoices, PDF/Word invoices, e-invoices |
| | Perceived Information Security | How secured do you think your current e-Invoicing system is with regards to sensitive customer information? | Likert Scale of 1-5 (Not secure at all - Very secure) |
| What kind of requirements are there for automation? | Requirements | As a customer what additional services would you like to have in future? | List of possible automated e-invoicing related functions (Option to write other is included) |
| | | As a customer how much would you like to have your e-invoicing process automated? | Likert Scale of 1-5 (Fully Manual - Fully Automated (touchless e-invoicing)) |

## Appendix 5 – Link to Collected Data and Analysis Materials

https://drive.google.com/drive/folders/1IhbXlvpgSVFEQzUki4-KFWfyQaP1ROWJ?usp=sharing

# Acknowledgement

I take pleasure in thanking my supervisor Ingrid Pappel for guiding me throughout the research study process. The opinions and feedback were extremely valuable.

Secondly, I would like to give thanks to my friend Viraj Brian Wijesuriya, a graduate student at the University of Oxford for brainstorming with me regarding emerging technologies as a recommendation during the research study process, who provided valuable insights.

Next, I would like to appreciate all the staff of e-Governance department for being supportive of my studies. Also, the opportunity to study and research the field of e-governance and add value to my career.

I am grateful for those who helped with answering the survey and taking the time to sit through interviews providing valuable aspects to this study.

Finally, thank you to my family back home, encouraging me throughout the study period.